\documentclass[%
 reprint,
superscriptaddress,
showpacs,preprintnumbers,
nofootinbib,
 amsmath,amssymb,
 aps,
]{revtex4-1}

\usepackage{subcaption}
\usepackage{soul}
\usepackage{xcolor}
\usepackage{natbib}
\usepackage{multirow}
\usepackage{float}
\usepackage{lipsum}
\usepackage{graphicx}
\usepackage{dcolumn}
\usepackage{bm}

\makeatletter
\newcommand*{\rom}[1]{\expandafter\@slowromancap\romannumeral #1@}
\makeatother

\usepackage[colorlinks=true, linkcolor=red, citecolor=blue, urlcolor=blue]{hyperref}

\begin{document}


\title{Modeling the spectrum and composition of ultrahigh-energy cosmic rays with two populations of extragalactic sources}

\author{Saikat Das}
\email{saikatdas@rri.res.in}
\affiliation{Astronomy \& Astrophysics Group, Raman Research Institute, Bengaluru 560080, India}

\author{Soebur Razzaque}
\email{srazzaque@uj.ac.za}
\affiliation{Centre for Astro-Particle Physics (CAPP) and Department of Physics, University of Johannesburg, PO Box 524, Auckland Park 2006, South Africa}

\author{Nayantara Gupta}
\email{nayan@rri.res.in}
\affiliation{Astronomy \& Astrophysics Group, Raman Research Institute, Bengaluru 560080, India}





\date{\today}

\begin{abstract}
We fit the ultrahigh-energy cosmic-ray (UHECR, $E\gtrsim0.1$ EeV) spectrum and composition data from the Pierre Auger Observatory at energies $E\gtrsim5\cdot10^{18}$ eV, i.e., beyond the ankle using two populations of astrophysical sources. One population, accelerating dominantly protons ($^1$H), extends up to the highest observed energies with maximum energy close to the GZK cutoff and injection spectral index near the Fermi acceleration model; while another population accelerates light-to-heavy nuclei ($^4$He, $^{14}$N, $^{28}$Si, $^{56}$Fe) with a relatively low rigidity cutoff and hard injection spectrum. A significant improvement in the combined fit is noted as we go from a one-population to two-population model. For the latter, we constrain the maximum allowed proton fraction at the highest-energy bin within 3.5$\sigma$ statistical significance. In the single-population model, low-luminosity gamma-ray bursts turn out to match the best-fit evolution parameter. In the two-population model, the active galactic nuclei is consistent with the best-fit redshift evolution parameter of the pure proton-emitting sources, while the tidal disruption events could be responsible for emitting heavier nuclei. We also compute expected cosmogenic neutrino flux in such a hybrid source population scenario and discuss possibilities to detect these neutrinos by upcoming detectors to shed light on the sources of UHECRs.

\end{abstract}

\maketitle


\section{\label{sec:introduction}Introduction\protect}

Identifying the sources of ultrahigh-energy cosmic rays (UHECRs, $E\gtrsim0.1$ EeV) is one of the outstanding problems in astroparticle physics \citep{Kotera_2011, Anchordoqui_2019}. Active Galactic Nuclei (AGNs) residing at the centers of nearby radio-galaxies are considered to be a potential candidate source class of UHECR acceleration \citep{Rachen_1993a, Rachen_1993b, Dermer_2009, Wang_2017, Eichmann_2018}. Studies involving the origin of TeV $\gamma$-rays assert blazars as ideal cosmic accelerators \citep{Essey_2010, Razzaque_2012, Murase_2012, Resconi_2017}. A recent analysis by the Pierre Auger Observatory has found a possible correlation between starburst galaxies and the observed intermediate scale anisotropy in UHECR arrival directions, with a statistical significance of 4$\sigma$ in contrast to isotropy \citep{Anchordoqui_1999, Attallah_2018, PAO_2018ApJ}. There are also propositions of other transient high-energy phenomena like gamma-ray bursts (GRBs) \citep{Waxman_1995, Vietri_1995, Dermer_2010, Globus_2015, Biehl_2018b, Zhang_2018}, tidal disruption events (TDEs) of white dwarfs or neutron stars \citep{Farrar_2014, Batista_2017, Biehl_2018a, Guepin_2018}, as well as, pulsar winds \citep{Lemoine_2015, Kotera_2015} which can reach the energy and flux required to explain the observed UHECR spectrum. Nevertheless, a direct correlation of these known source catalogs, derived from X-ray and $\gamma$-ray observations, with an observed UHECR event is yet to be made \citep{Elbert_1995, Farrar_1998, Virmani_2002, Gorbunov_2004}. The different source classes allow an extensively wide range of UHECR parameters to be viable in the acceleration region. UHECRs produce neutrinos and $\gamma$-rays on interactions with the cosmic background photons during their propagation over cosmological distances. The current multimessenger data can only constrain UHECR source models and provide hints towards plausible accelerator environments \citep{Batista_2019, Heinze_2019}, rejecting the possibility of a pure proton composition at the highest energies \citep{Berezinsky_2011, Gelmini_2012, Fang_2016, Supanitsky_2016, Heinze_2016}. Deflections in Galactic and extragalactic magnetic fields pose an additional challenge in UHECR source identification. 

The Pierre Auger Observatory (PAO) in Malarg\"ue, Argentina \citep{PAO_2015} and the Telescope Array (TA) experiment in Utah, United States \citep{Abbasi_2016} are attaining unprecedented precision in the measurement of UHECR flux, composition, and arrival directions from 0.3 EeV to beyond 100 EeV using their hybrid detection technique \citep{PAO_2017, Abbasi_2018b}. On incidence at the Earth's atmosphere, these energetic UHECR nuclei initiate hadronic cascades which are intercepted by the surface detector (SD), and the simultaneous fluorescence light emitted by the Nitrogen molecules in the atmosphere is observed using the fluorescence detector (FD). This extensive air shower (EAS) triggered by the UHECRs is recorded to measure the maximum shower-depth distribution ($X_{\rm max}$) \citep{Abbasi_2018a}. However, even with the large event statistics observed by PAO, the mass composition is not as well constrained as the spectrum and anisotropy up to $\sim100$ EeV \citep{PAO_2017JCAP}. The first two moments of $X_{\rm max}$, viz., the mean $\langle X_{\rm max}\rangle$, and its fluctuation from shower-to-shower $\sigma(X_{\rm max})$ serves the purpose of deducing the mass composition. The standard shower propagation codes, eg., \textsc{corsika} \citep{CORSIKA}, \textsc{conex} \citep{Pierog_2006}, etc., depend on the choice of a hadronic interaction model and photodisintegration cross-section, which are extrapolations of the hadronic physics to the ultrahigh-energy regime. Uncertainties in these models propagate to uncertainties in the reconstruction of the mass-composition of observed events. Lifting the degeneracy in the mass composition will be essential to constrain the source models.

The current LHC-tuned hadronic interaction models viz., \textsc{sybill2.3c} \citep{Riehn_2015}, \textsc{epos-lhc} \citep{Pierog_2015}, and \textsc{qgsjet-II.04} \citep{Ostapchenko_2011} differ in their inherent assumptions and thus lead to different inferences of the mass composition using the same observed data. Current estimates from PAO predict that the relative fraction of protons decreases with increasing energy above $10^{18.3}$ eV for all three models. For the first two models, N dominates at $10^{19.6}$ eV, while for the third model, the entire contribution at the highest energy comes from He. The ankle at $E\approx10^{18.7}$ eV corresponds to a mixed composition with He dominance and lesser contributions from N and H, except for \textsc{qgsjet-II.04} which suggests a zero N fraction \citep{PAO_comp2017}. The ankle is often inferred as a transition between two or more different populations of sources, leading to a tension between the preference of Galactic or extragalactic nature of the sub-ankle spectrum. Based on the observed anisotropy and light composition, some UHECR models invoke increased photohadronic interactions of UHECRs in the environment surrounding the source. The magnetic field of the surrounding environment can confine the heavier nuclei with energies higher than that corresponding to the ankle, while they undergo photo-disintegration/spallation to produce the light component in the sub-ankle region \citep{Unger_2015, Kachelriess_2017, Supanitsky_2018}. This requires only a single class of UHECR sources that accelerate protons and nuclei. However, it is also possible to add a distinct light nuclei population of extragalactic origin that can explain the origin of the sub-ankle spectrum \citep{Wang_2007, Aloisio_2014, Muzio_2019}. A purely protonic component, in addition to a Milky Way-like nuclear composition, has also been studied \citep{Muzio_2019}. The proton fraction in the UHECR spectrum for various source models can be constrained through composition studies and compliance to multimessenger data \citep{Moller_2019, Vliet_2019}.

In this work, first, we perform a combined fit of spectrum and composition data at $E \gtrsim 5\cdot 10^{18}$ eV measured by PAO \citep{PAO_ICRC_2017}, to find the best-fit parameters for a single-population of extragalactic UHECR sources injecting a mixed composition of representative elements ($^1$H, $^4$He, $^{14}$N, $^{28}$Si, $^{56}$Fe). The best-fit $^1$H abundance fraction is found to be zero in this case, conceivable within our choice of the photon background model, photodisintegration cross-section, and hadronic interaction model. Next, we show that within the permissible limit of current multimessenger photon and neutrino flux upper limits \citep{Ackermann_2015, Aartsen_2018}, the addition of a purely protonic ($^1$H) component up to the highest-energy bin can significantly improve the combined fit of spectrum and composition. We consider this component originates from a separate source population than the one accelerating light-to-heavy nuclei and fit the region of the spectrum above the ankle, i.e., $E\gtrsim 5\cdot 10^{18}$ eV. The best-fit values of the UHECR parameters are calculated for both the populations, allowing for a one-to-one comparison with the single-population case. We study the effect of variation of the proton injection spectral index, which is not done in earlier studies and indicate the maximum allowed proton fraction at the highest-energy bin up to 3.5$\sigma$ statistical significance. We calculate the fluxes of cosmogenic neutrinos that can be produced by these two populations. We also explore the prospects of their observation by upcoming detectors, and probe the proton fraction at the highest-energy of the UHECR spectrum. Lastly, we take into account the redshift evolution of the two source populations, which is found to further improve the combined fit. We interpret the credibility of the best-fit redshift distributions in light of known candidate classes.

We explain our model assumptions and simulation setup in Sec.~\ref{sec:framework} and present our results for both single-population and two-population models in Sec.~\ref{sec:result}. We discuss our results and possible source classes in light of the two-population model in Sec.~\ref{sec:discussions} and draw our conclusions in Sec.~\ref{sec:conclusions}.

\section{\label{sec:framework}UHECR propagation and shower depth distribution\protect}

UHECRs propagate over cosmological distances undergoing a variety of photohadronic interactions. These interactions lead to the production of secondary particles, viz., cosmogenic neutrinos and photons. The dominant photopion production of UHECR protons on the cosmic microwave background (CMB) via delta resonance occurs at $\approx6.8\times10^{19}$ eV, producing neutral and charged pions ($\pi^0$, $\pi^+$) with 2/3 and 1/3 probability, respectively. The neutral pions decay to produce $\gamma$-rays ($\pi^0\rightarrow\gamma\gamma$), while the charged pions decay to produce neutrinos ($\pi^+\rightarrow\mu^+ + \nu_\mu \rightarrow\mathrm{e^+} + \nu_e + \overline{\nu}_\mu + \nu_\mu$). Neutrinos can also be produced through other $p\gamma$ processes and neutron beta decay ($n\rightarrow p + \mathrm{e^-} + \overline{\nu}_e$). Bethe-Heitler interaction of UHECR protons of energy $\approx4.8\times10^{17}$ eV with CMB photons can produce $\mathrm{e^+e^-}$ pairs. The $\mathrm{e^+}$ and $\mathrm{e^-}$ produced through various channels can iteratively produce high-energy photons by inverse-Compton scattering of cosmic background photons or synchrotron radiation in the extragalactic magnetic field (EGMF). The produced photons can undergo Breit-Wheeler pair production. All these interactions also hold for heavier nuclei ($^A_ZX$, $Z>1$), in addition to photodisintegration. The interactions may also occur with the extragalactic background light (EBL), having energy higher than the CMB, with cosmic-rays of lower energy. Besides, all particles lose energy due to the adiabatic expansion of the universe. We consider $\Lambda$CDM cosmology with the parameter values $H_0=67.3$ km s$^{-1}$ Mpc$^{-1}$, $\Omega_m=0.315$, $\Omega_\Lambda=1-\Omega_m$ \cite{Olive_2014}. While cosmic rays are deflected by the Galactic and extragalactic magnetic fields, the neutrinos travel unaffected by matter or radiation fields, and undeflected by magnetic fields.

The observed spectrum depends heavily on the choice of injection spectrum. We consider all elements are injected by the source following the spectrum given by,
\begin{equation}
\dfrac{dN}{dE} = A_0 \sum_i K_i \bigg(\dfrac{E}{E_0}\bigg)^{-\alpha} f_{\rm cut}(E, ZR_{\rm cut}) \label{eq:injection}
\end{equation}
This represents an exponential cutoff power-law function, where $K_i$ and $\alpha$ are the abundance fraction of elements and spectral index at injection. $A_0$ and $E_0$ are arbitrary normalization flux and reference energy, respectively. A similar spectrum has been considered in the combined fit analysis by the PAO \citep{PAO_2017JCAP}. The broken exponential cutoff function is written as,
\begin{align}
f_{\rm cut} = 
\begin{cases}
1 & (E\leqslant ZR_{\rm cut})\\ \exp\bigg(1-\dfrac{E}{ZR_{\rm cut}}\bigg) & (E> ZR_{\rm cut})
\end{cases}
\end{align}

We use the \textsc{CRPropa 3} simulation framework to find the particle yields obtained at Earth after propagating over extragalactic space from the source to the observer \citep{CRPropa3}. We find the best-fit values of the UHECR parameters $\alpha$, rigidity cutoff ($R_{\rm cut}$) and $K_i$ for both one-population and two-population models. The normalization depends on the source model and the source population. The spectrum of EBL photons and its evolution with redshift is not as well known as for CMB. We use a latest and updated EBL model by Gilmore \textit{et al.} \citep{Gilmore_2012} and \textsc{talys} 1.8 photodisintegration cross-section \citep{Koning_2005}.

We use the parametrizations given by PAO based on the Heitler model of EAS to calculate the mean depth of cosmic-ray air shower maximum $\langle X_{\rm max}\rangle$ and its dispersion from the first two moments of $\ln A$  \citep{Matthews_2005, Abreu_2013}.
\begin{align}
\langle X_{\rm max} \rangle &= \langle X_{\rm max} \rangle_p + f_E \langle \ln A \rangle \\
\sigma^2 (X_{\rm max}) &= \langle \sigma^2_{\rm sh} \rangle + f_E^2 \sigma^2_{\ln A}
\end{align}
where $\langle X_{\rm max} \rangle_p$ is the mean maximum depth of proton showers and $f_E$ is a parameter which depends on the energy of the UHECR event,
\begin{equation}
f_E = \xi - \dfrac{D}{\ln 10} + \delta\log_{10}\bigg(\dfrac{E}{E_0}\bigg)
\end{equation} 
where $\xi$, $D$, and $\delta$ depend on the specific hadronic interaction model. $\sigma^2_{\ln A}$ is the variance of $\ln A$ distribution and $\langle \sigma_{\rm sh}^2 \rangle$ is the average variance of $X_{\rm max}$ weighted according to the $\ln A$ distribution,
\begin{equation}
\langle \sigma_{\rm sh}^2 \rangle = \sigma_p^2[1+ a \langle \ln A \rangle + b \langle (\ln A)^2 \rangle]
\end{equation}
where $\sigma_p^2$ is the $X_{\rm max}$ variance for proton showers depending on energy and three model-dependent parameters. In this work, we use the updated parameter values\footnote{S. Petrera and F. Salamida (2018), Pierre Auger Observatory} obtained from the \textsc{conex} simulations \citep{Pierog_2006}, for one of the post-LHC hadronic interaction models, \textsc{sybill2.3c}.

\section{\label{sec:result}Results\protect}

\begin{figure*}[hbt]
\centering
\begin{subfigure}{.5\textwidth}
  \centering
  \includegraphics[width=\textwidth]{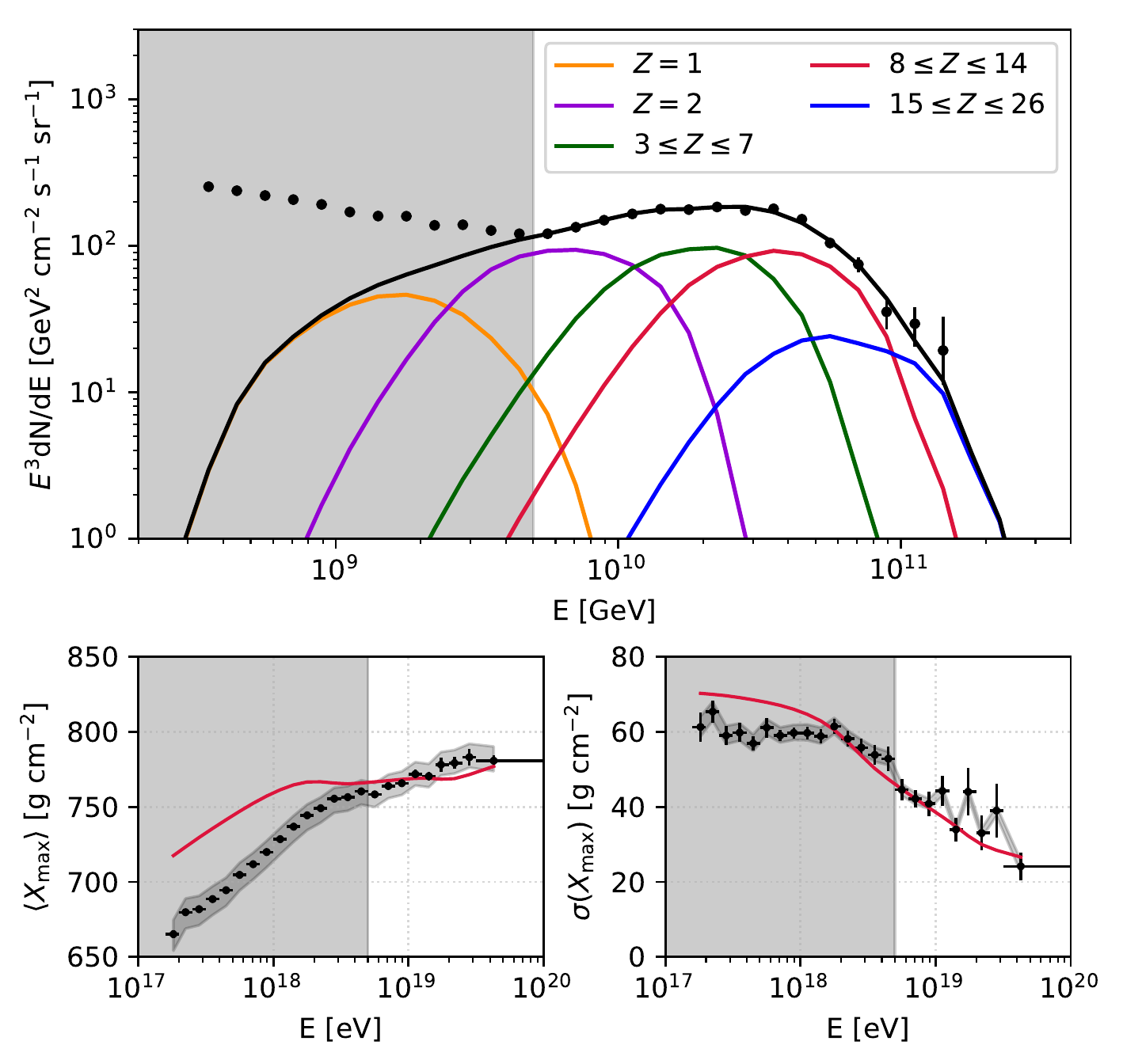}
  \caption{One-population model}
\end{subfigure}%
\begin{subfigure}{.5\textwidth}
  \centering
  \includegraphics[width=\textwidth]{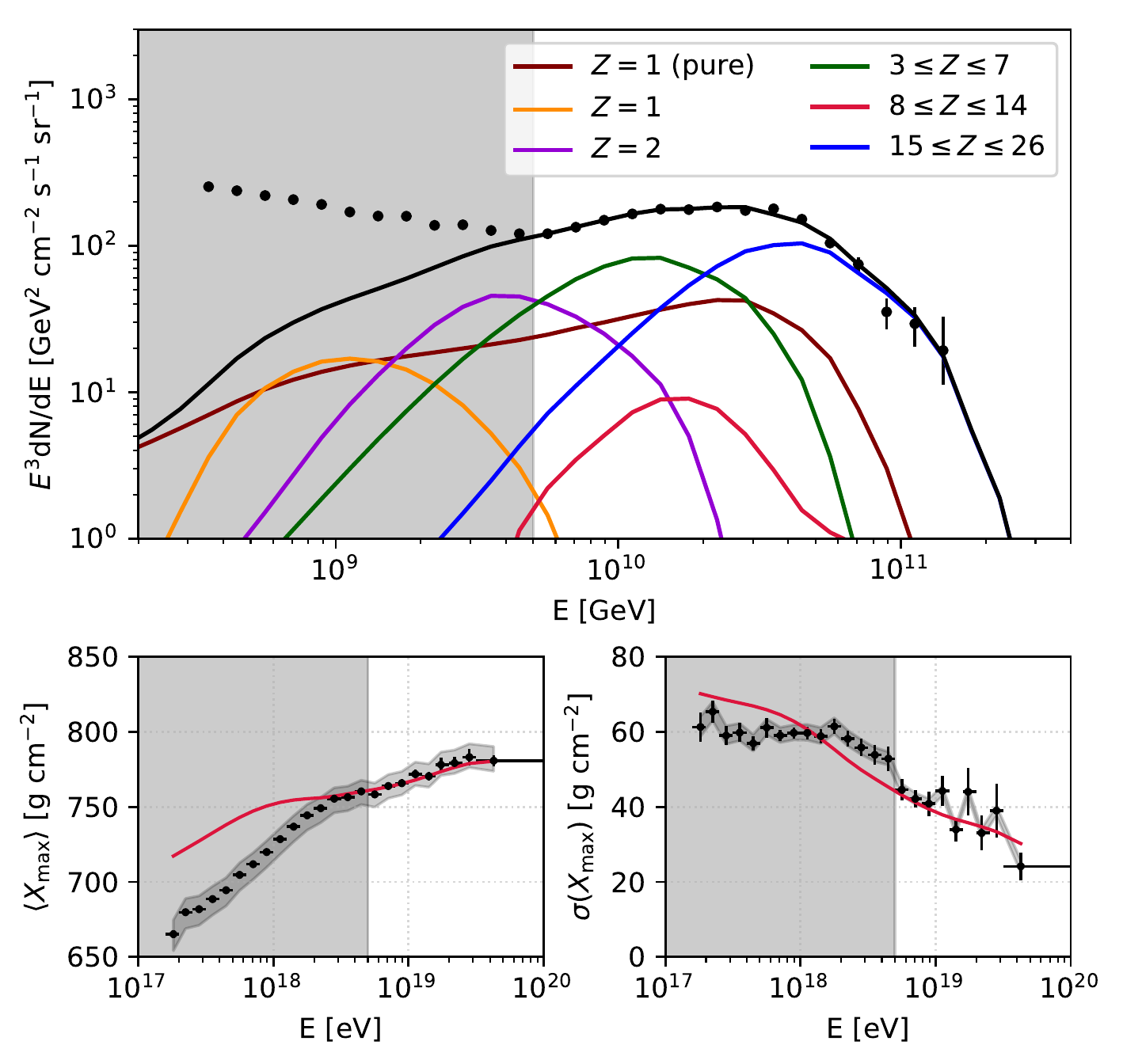}
  \caption{Two-population model ($\alpha_1=2.2$)}
\end{subfigure}
\begin{subfigure}{.5\textwidth}
  \centering
  \includegraphics[width=\textwidth]{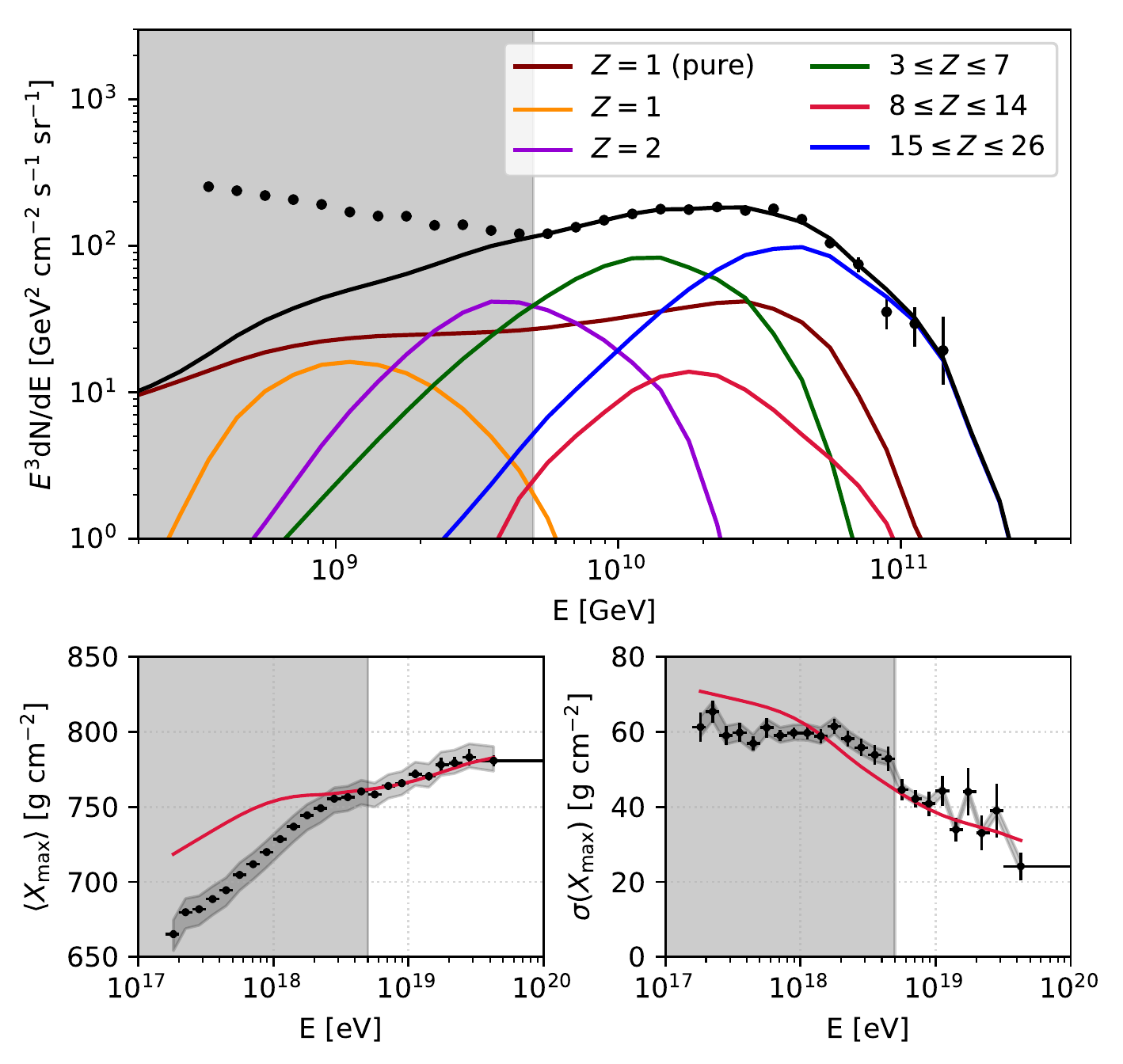}
  \caption{Two-population model ($\alpha_1=2.4$)}
\end{subfigure}%
\begin{subfigure}{.5\textwidth}
  \centering
  \includegraphics[width=\textwidth]{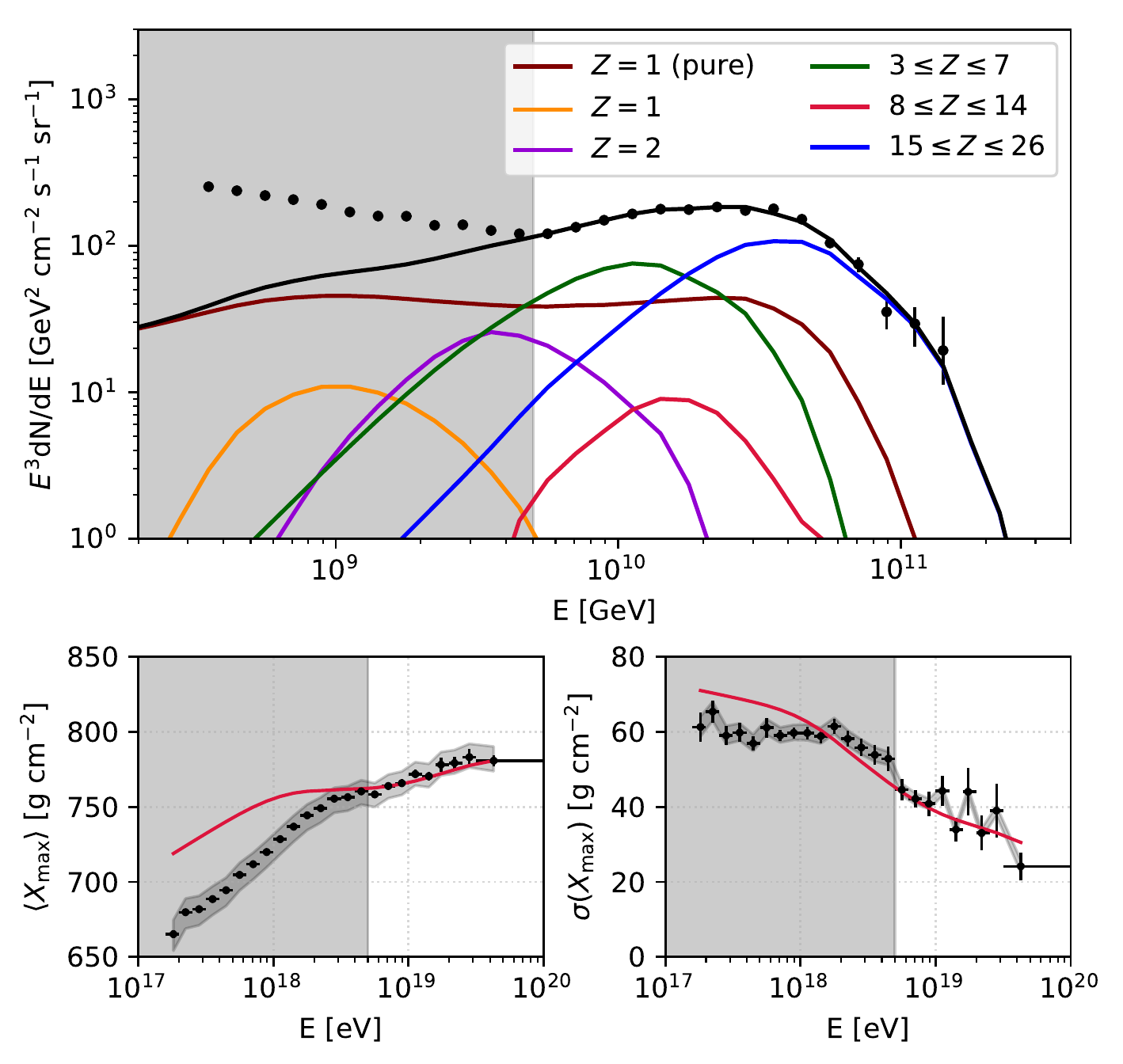}
  \caption{Two-population model ($\alpha_1=2.6$)}
\end{subfigure}
\caption{\small{UHECR spectrum and composition for the best-fit parameters of single-population and two-population models in the flat ($m=0$) cosmological evolution scenario. For the latter case, the resulting spectra for different injection spectral index of the pure-proton component are shown.}}
\label{fig:spec_CTG}
\end{figure*}
\begin{figure*}[hbt]
\centering
\begin{subfigure}{.5\textwidth}
  \centering
  \includegraphics[width=\textwidth]{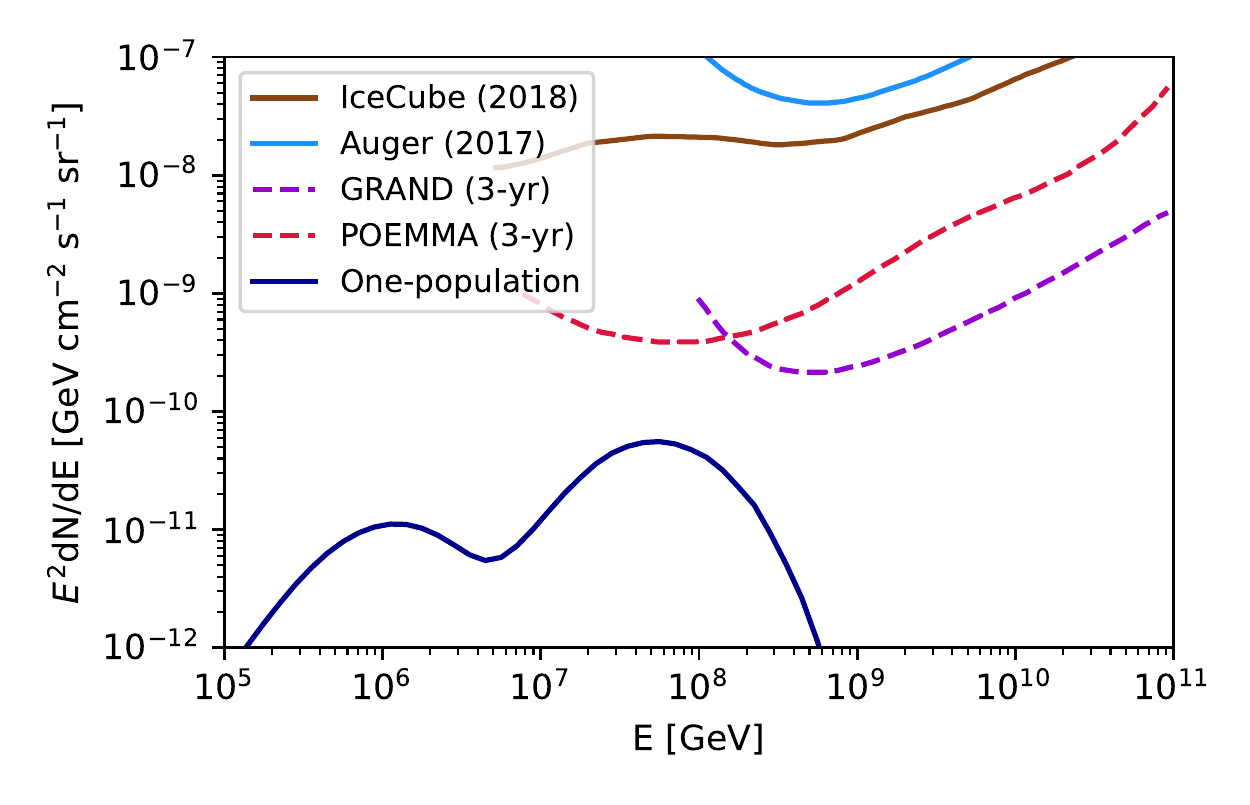}
  \caption{One-population model}
\end{subfigure}%
\begin{subfigure}{.5\textwidth}
  \centering
  \includegraphics[width=\textwidth]{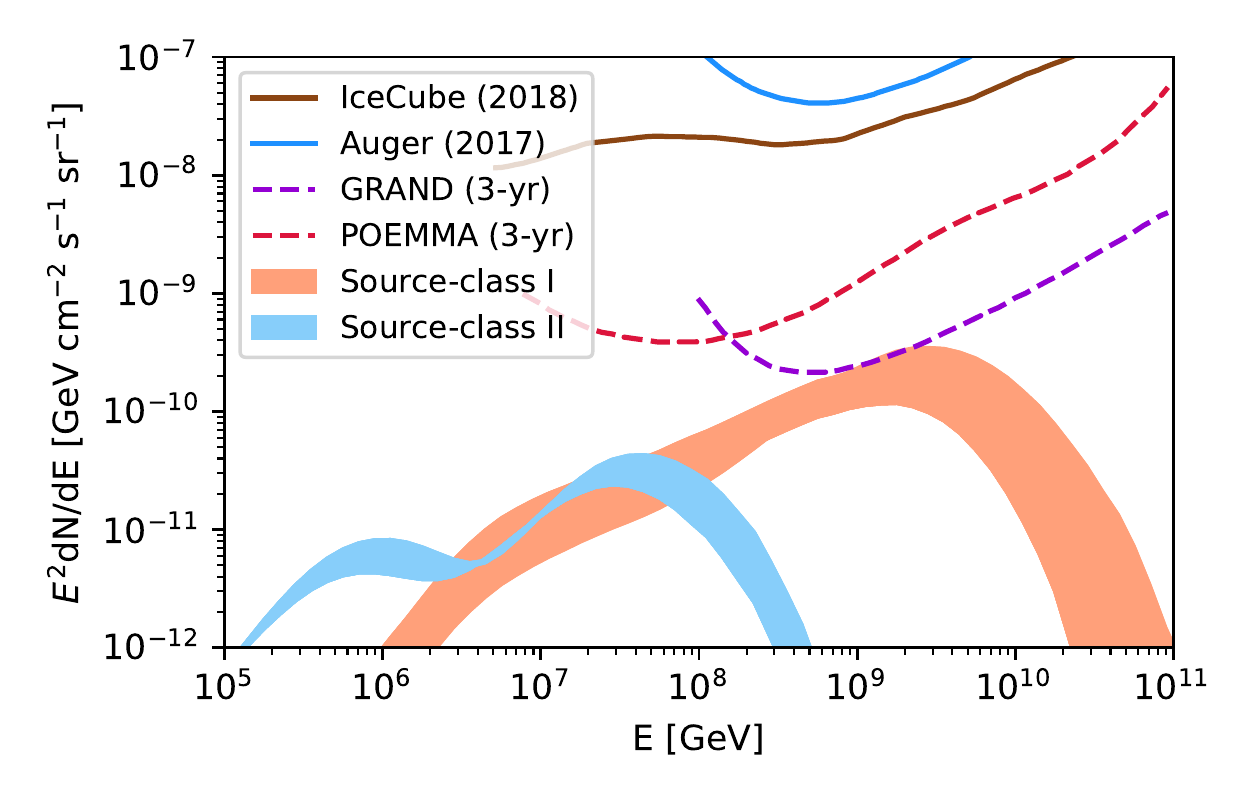}
  \caption{Two-population model ($\alpha_1=2.2$)}
\end{subfigure}
\begin{subfigure}{.5\textwidth}
  \centering
  \includegraphics[width=\textwidth]{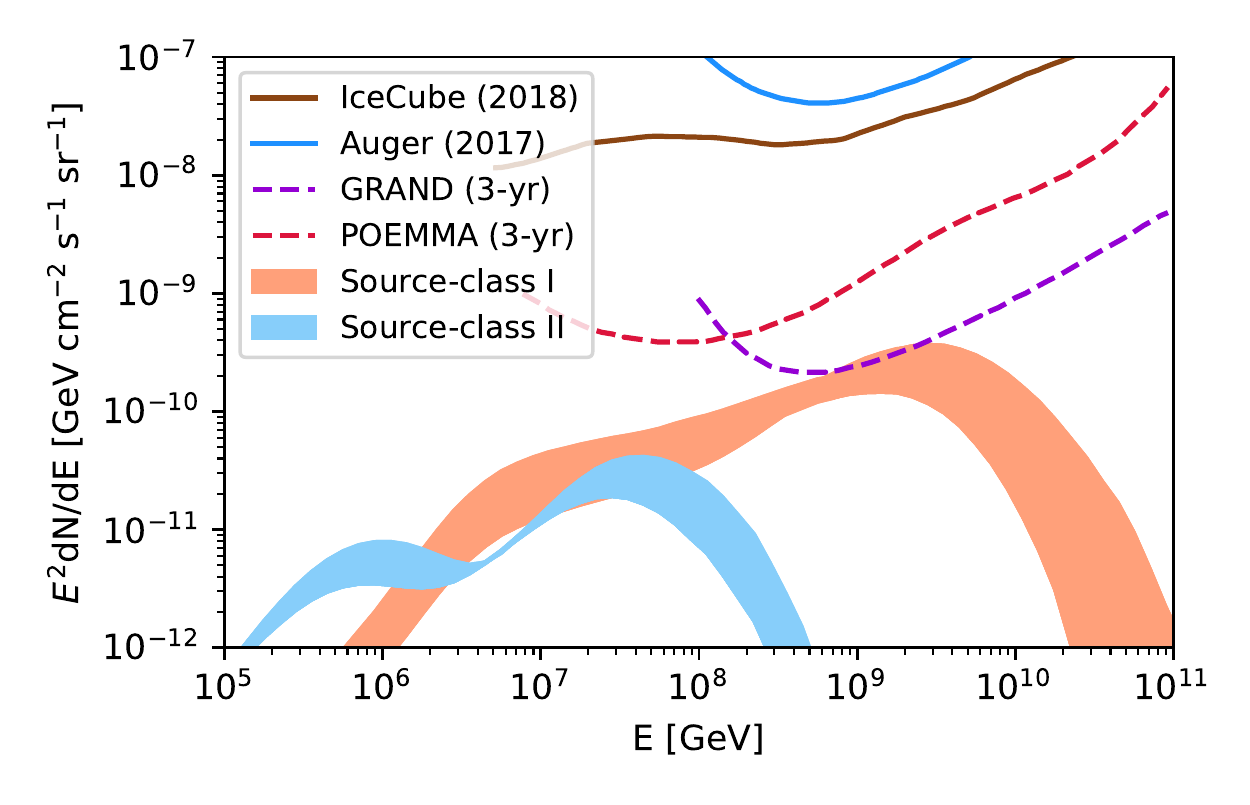}
  \caption{Two-population model ($\alpha_1=2.4$)}
\end{subfigure}%
\begin{subfigure}{.5\textwidth}
  \centering
  \includegraphics[width=\textwidth]{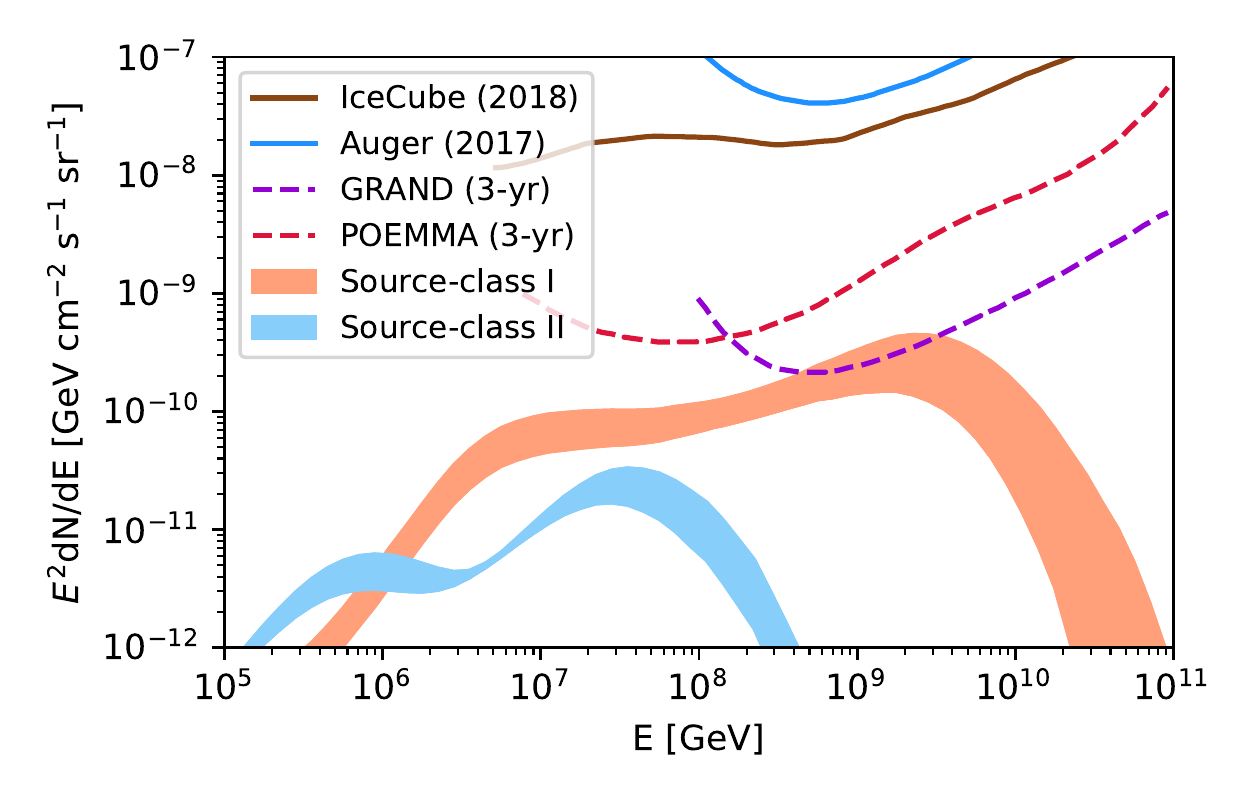}
  \caption{Two-population model ($\alpha_1=2.6$)}
\end{subfigure}
\caption{\small{The all-flavor cosmogenic neutrino fluxes for one-population and two-population models (without cosmological evolutions) along with the sensitivity of currently operating and future neutrino detectors. The neutrino flux originating from distinct source populations up to $f_{\rm H}=20.0$\% are shown for proton injection index $\alpha_1 = 2.2$, 2.4, and 2.6 in the top-right, bottom-left and bottom-right panels.}}
\label{fig:neu_CTG}
\end{figure*}

We perform a combined fit of our UHECR source models to the spectrum and composition data measured by PAO \citep{PAO_spec2017, PAO_comp2017}, for one-population and two-population model of the UHECR sources. The fit region corresponds to energies above the ankle, i.e., $E\gtrsim 5\cdot10^{18}$ eV in the spectrum, as well as, composition. We calculate the goodness-of-fit using the standard $\chi^2$ formalism,
\begin{equation}
\chi^2_j = \sum_{i=1}^N \bigg[\dfrac{y_i^{\rm obs}(E) - y_i^{\rm mod}(E; a_M)}{\sigma_i}\bigg]^2
\end{equation}
where the subscript $j$ corresponds to any of the three observables, viz., spectrum, $X_{\rm max}$, or $\sigma(X_{\rm max})$. To find the best-fit cases, we minimize the sum of all the $\chi^2_j$ values. Here $y_i^{\rm obs}(E)$ is the measured value of an observable in the $i-$th energy bin corresponding to a mean energy $E$ and $y_i^{\rm mod}(E; a_M)$ is the value obtained numerically. $a_M$ are the best-fit values of $M$ parameters varied in the simulations. $\sigma_i$ are the errors provided by PAO. We denote the spectral fit as $\chi^2_{\rm spec}$ and the composition fit as $\chi^2_{\rm comp}$. The latter represents the goodness-of-fit considering $X_{\rm max}$ and $\sigma({X_{\rm max}})$ simultaneously. In the following subsections, we demonstrate the one-population model in Subsec.~\ref{subsec:one-pop}, the transition due to the addition of an exclusive proton injecting class in Subsec.~\ref{subsec:two-pop}, and finally the effects of redshift distribution in Subsec.~\ref{subsec:z_evol}.
\subsection{\label{subsec:one-pop}One-population model}
\begin{table}
\caption{\label{tab:one-pop}UHECR best-fit parameter set for the one-population model (flat evolution, $m=0$)}
\centering
\begin{tabular}{lllll}
\hline
\textbf{Parameter} & \multicolumn{3}{l}{\textbf{Description}} & \textbf{Values}\\ 
\hline
$\alpha$ & \multicolumn{3}{l}{Source spectral index} & -0.7 \\ 
$\log_{10}(R_{\rm cut}/\rm V)$ & \multicolumn{3}{l}{Cutoff rigidity} & 18.2 EV\\
$z_{\text{max}}$ & \multicolumn{3}{l}{Cutoff redshift} & 1.0 (fixed)\\
$m$ & \multicolumn{3}{l}{Source evolution index} & 0.0 (fixed)\\
\hline
$K_i$(\%) & \hspace*{-0.1cm} H & \hspace*{0.3cm} He & \hspace*{0.3cm} N & \hspace*{-0.4cm} Si \hspace*{0.8cm} Fe\\
 & \hspace*{-0.1cm} 0.0 & \hspace*{0.3cm} 95.6 & \hspace*{0.3cm} 4.1 & \hspace*{-0.4cm} 0.3 \hspace*{0.5cm} 0.0073\\
 \hline
 $\chi^2_{\rm tot}/$d.o.f & $\chi^2_{\rm spec}$ & $\chi^2_{\rm comp}$ & &\\
 56.19/25 & 9.94 & 46.25 & &\\
 \hline
\end{tabular}
\end{table}
We start by considering a single population of extragalactic sources up to a redshift $z=1$, injecting a mixed composition of representative elements $^1$H, $^4$He, $^{14}$N, $^{28}$Si, and $^{56}$Fe following an injection spectrum given by Eq.~\ref{eq:injection}. The elements are injected with energy between $0.1-1000$ EeV. The combined fit analysis done by PAO argues that only particles originating from $z\lesssim 0.5$ are able to reach Earth with $E>5\cdot 10^{18}$ eV \citep{Greisen_1966, Zatsepin_1966, PAO_2017JCAP}. Indeed, in our case, the contribution at the spectral cutoff comes from $^{56}$Fe. Hence, the sources which are located further in the distance than $z_{\rm max}=1$ are unable to contribute to the spectrum above the ankle ($\approx10^{18.7}$ eV) \citep[see, eg., Appendix C of][]{Batista_2019}. This is because, as the distance of such heavy nuclei injecting sources increases, the rate of photodisintegration also gradually increases, thus decreasing their survival rate at the highest energies. Moreover, it was found that increasing $z_{\rm max}$ has no effect on the best-fit parameters found with $z_{\rm max}=1$ \citep{Heinze_2019}. The source distribution is assumed to be uniform over comoving distance. In a later subsection, we check the effects of a non-trivial redshift evolution for one-population model.

We scan the parameter space by varying the rigidity cutoff $\log_{10}(R_{\rm cut}/\rm V)$ between [18.0, 18.5] with a grid spacing of 0.1 and the injection spectral index $\alpha$ between [-1.5, 1.0] with a grid spacing of 0.1. For each set of values \{$\alpha$, $\log_{10}(R_{\rm cut}/\rm V)$\}, we find the best-fit abundance fraction of the injected elements. The number of physical parameters varied is 7 and we consider the normalization to be an additional free parameter. Hence the number of degrees of freedom (d.o.f) is $N_{\rm d}= 33-7-1=25$ in this model, since the fitting is done to a total of 33 data points. All the parameter values for the best-fit case of the single-population model are listed in Table~\ref{tab:one-pop}.


\renewcommand{\arraystretch}{1.0}
\begin{table*}[hbt]
 \caption{\label{tab:two-pop} Best-fits to UHECR spectrum and composition for two-population model without cosmological evolution}
  \begin{ruledtabular}
\centering
 \begin{tabular}{ccc|cccccc|ccc|l}
\multicolumn{3}{c|}{\texttt{\textbf{Source-class I}}} & \multicolumn{6}{c|}{\textbf{\texttt{Source-class II}}} & \multicolumn{4}{c}{\textbf{Goodness-of-fit}}\\ [0.15cm]
\hline
$\alpha_{1}$ & $f_{\rm H}$(\%) & $\log_{10}(R_{\rm cut, 1}/\rm V)$ & $\alpha_{2}$ & $\log_{10}(R_{\rm cut, 2}/\rm V)$ & $K_{He}$ & $K_{N}$ & $K_{Si}$ & $K_{Fe}$ & $\chi^2_{\rm spec}$ & $\chi^2_{\rm comp}$ & $\chi^2_{\rm tot}$ & ID \#\\ [0.15cm]
\hline
\textbf{2.2} & 1.0\% & 19.5 & 0.6 & 18.30 & 74.75 & 22.50 & 2.00 & 0.75 & 16.97 & 23.03 & 40.00 & \rom{1}\\
& 1.5\% & 19.5 & 0.9 & 18.30 & 53.00 & 44.25 & 0.00 & 2.75 & 14.68 & 15.52 & \textbf{30.20} & \rom{2}\\
& 2.0\% & 19.5 & 1.2 & 18.30 & 41.50 & 52.50 & 0.00 & 6.00 & 17.03 & 15.38 & 32.41 & \rom{3}\\
& 2.5\% & 19.6 & 0.6 & 18.30 & 73.25 & 24.25 & 1.75 & 0.75 & 13.86 & 21.68 & 35.54 & \rom{4}\\
& 5.0\% & 19.7 & 0.5 & 18.28 & 76.50 & 21.25 & 1.75 & 0.50 & 12.11 & 25.56 & 37.67 & \rom{5}\\
& 7.5\% & 19.8 & 0.3 & 18.28 & 82.25 & 16.25 & 1.25 & 0.25 & 13.47 & 28.00 & 41.47 & \rom{6}\\
& 10.0\% & 19.8 & 0.6 & 18.28 & 71.25 & 26.50 & 1.50 & 0.75 & 14.07 & 28.74 & 42.81 & \rom{7}\\ 
& 12.5\% & 19.9 & 0.3 & 18.26 & 82.50 & 16.00 & 1.25 & 0.25 & 14.28 & 29.57 & 43.85 & \rom{8}\\
& 15.0\% & 20.0 & 0.3 & 18.28 & 81.75 & 16.75 & 1.25 & 0.25 & 16.62 & 29.31 & 45.93 & \rom{9}\\
& 17.5\% & 20.0 & 0.3 & 18.26 & 82.25 & 16.25 & 1.25 & 0.25 & 15.85 & 30.51 & 46.36 & \rom{10}\\
& 20.0\% & 20.1 & 0.3 & 18.28 & 81.50 & 17.00 & 1.25 & 0.25 & 17.65 & 30.10 & 47.75 & \rom{11}\\[0.15cm]
\hline
\textbf{2.4} & 1.0\% & 19.5 & 0.8 & 18.28 & 56.75 & 39.75 & 1.25 & 2.25 & 14.86 & 20.46 & 35.32 & \rom{12}\\
& 1.5\% & 19.5 & 1.3 & 18.30 & 18.75 & 70.25 & 0.00 & 11.00 & 21.15 & 15.48 & 36.63 & \rom{13}\\
& 2.0\% & 19.6 & 0.6 & 18.28 & 68.75 & 28.75 & 1.50 & 1.00 & 13.55 & 21.80 & 35.35 & \rom{14}\\
& 2.5\% & 19.6 & 0.9 & 18.30 & 45.25 & 51.00 & 0.75 & 3.00 & 12.60 & 18.13 & \textbf{30.73} & \rom{15}\\
& 5.0\% & 19.7 & 0.8 & 18.28 & 54.50 & 42.25 & 1.25 & 2.00 & 12.13 & 22.16 & 34.39 & \rom{16}\\
& 7.5\% & 19.8 & 0.6 & 18.28 & 71.00 & 26.00 & 2.25 & 0.75 & 12.36 & 27.10 & 39.46 & \rom{17}\\
& 10.0\% & 19.9 & 0.5 & 18.28 & 75.75 & 21.75 & 2.00 & 0.50 & 13.78 & 28.42 & 42.20 & \rom{18}\\
& 12.5\% & 19.9 & 0.6 & 18.26 & 71.50 & 25.50 & 2.25 & 0.75 & 12.99 & 30.22 & 43.21 & \rom{19}\\
& 15.0\% & 20.0 & 0.5 & 18.28 & 74.75 & 22.75 & 2.00 & 0.50 & 14.93 & 29.32 & 44.25 & \rom{20}\\
& 17.5\% & 20.1 & 0.3 & 18.26 & 82.00 & 16.25 & 1.50 & 0.25 & 14.60 & 30.43 & 45.03 & \rom{21}\\ 
& 20.0\% & 20.2 & 0.3 & 18.28 & 80.50 & 17.75 & 1.50 & 0.25 & 16.40 & 29.69 & 46.09 & \rom{22}\\[0.15cm]
\hline
\textbf{2.6} & 1.0\% & 19.5 & 1.3 & 18.30 & 0.00 & 84.50 & 0.00 & 15.50 & 21.43 & 22.86 & 44.29 & \rom{23}\\
& 1.5\% & 19.6 & 0.8 & 18.30 & 46.25 & 49.50 & 1.75 & 2.50 & 14.57 & 23.97 & 38.54 & \rom{24}\\
& 2.0\% & 19.6 & 1.1 & 18.30 & 0.00 & 91.50 & 0.00 & 8.50 & 12.03 & 19.63 & \textbf{31.66} & \rom{25}\\
& 2.5\% & 19.6 & 1.3 & 18.30 & 0.00 & 83.50 & 3.00 & 13.50 & 18.18 & 22.07 & 40.25 & \rom{26}\\
& 5.0\% & 19.7 & 1.3 & 18.30 & 0.00 & 83.75 & 4.25 & 12.00 & 14.36 & 25.43 & 39.79 & \rom{27}\\
& 7.5\% & 19.8 & 1.1 & 18.30 & 0.00 & 90.75 & 2.00 & 7.25 & 12.27 &	24.77 & 37.04 & \rom{28}\\
& 10.0\% & 19.9 & 0.8 & 18.28 & 51.00 & 44.00 & 3.00 & 2.00 & 13.18 & 26.74 & 39.92 & \rom{29}\\
& 12.5\% & 19.9 & 1.0 & 18.28 & 20.00 & 71.75 & 3.50 & 4.75 & 13.35 & 29.01 & 42.36 & \rom{30}\\
& 15.0\% & 20.0 & 0.8 & 18.28 & 49.50 & 45.50 & 3.00 & 2.00 & 14.69 & 27.75 & 42.44 & \rom{31}\\
& 17.5\% & 20.1 & 0.6 & 18.28 & 61.75 & 35.00 & 2.25 & 1.00 & 16.88 & 27.57 & 44.45 & \rom{32}\\
& 20.0\% & 20.1 & 0.7 & 18.26 & 62.75 & 33.00 & 3.00 & 1.25 & 14.25 & 30.00 & 44.25 & \rom{33}\\
\end{tabular}
\end{ruledtabular}
 
\end{table*}

We see that the best-fit $^1$H fraction turns out to be zero, and a non-zero $^{56}$Fe component is unavoidable in this case. Indeed from the best-fit spectrum, shown in the upper left panel of Fig.~\ref{fig:spec_CTG}, the contribution from $Z=1$ component above $5\cdot10^{18}$ eV is infinitesimal. Since the heavier nuclei must come from nearby sources, for them to survive at the highest energies, the maximum rigidity, in this case, suggests that the cutoff in the spectrum originates from maximum acceleration energy at the sources. The fit, however, corresponds to a negative injection spectral index, which is difficult to explain by either the existing particle acceleration models or by sufficient hardening due to photohadronic interactions in the environment surrounding the source. The slope of the simulated $X_{\rm max}$ plot (cf. Fig.~\ref{fig:spec_CTG}), in comparison to data, suggests that the addition of a light element above $10^{19}$ eV can improve the fit. Motivated by these aforementioned characteristics of the combined fit, it is impulsive to add the contribution from another source population and check the effects on the spectrum and composition. 

\subsection{\label{subsec:two-pop}Two-population model}

We consider a discrete extragalactic source population injecting $^1$H following the spectrum of Eq.~\ref{eq:injection}. We refer to this as the \texttt{source-class I} (abbv. \texttt{Cls-I}). This pure-proton component has a distinct rigidity cutoff $R_{\rm cut,1}$, and injection spectral index $\alpha_1\gtrsim2$, such that the spectrum extends up to the highest-energy bin of the observed UHECR spectrum. The normalization $A_1 =A_p$ is fixed by the condition $J_p(E_h)=f_{\rm H}J(E_h)$, where $J(E)=dN/dE$ of the observed spectrum and $E_h$ is the mean energy of the highest-energy bin. $f_{\rm H}$ is an additional parameter that takes care of the proton fraction in the highest-energy bin of the UHECR spectrum. 

Another population (\texttt{source-class II}, abbv. \texttt{Cls-II}) injects light-to-heavy nuclei, viz., $^4$He, $^{14}$N, $^{28}$Si, and $^{56}$Fe, as we have already seen that for a mixed composition at injection, the contribution of $^1$H abundance tends to be zero above the ankle energy. \texttt{Cls-II} also follows the spectrum in Eq.~\ref{eq:injection} with rigidity cutoff $R_{\rm cut,2}$ and injection spectral index $\alpha_2$, and the abundance fraction at injection given by $K_{i}$ ($\sum_i K_i =100$\%). The normalization $A_2$ in this case is a free parameter which is adjusted to fit the spectrum and composition. As in the single-population model, here too, we set the maximum redshift of the sources to $z_{\rm max}=1$. Athough the anisotropy of UHECR arrival directions suggest that the observed spectrum depends on the position distribution of their sources, a definitive source evolution model is difficult to find. The rigidity cutoff and the injection spectral index will vary widely with the variation of evolution function and its exponent. We first consider that both the source populations are devoid of redshift evolution, i.e., $m=0$ in the $(1+z)^m$ type of source evolution models. Afterwards, in the next subsection, we present the $m\ne 0$ cases for one- and two-population models.

The cumulative contribution of \texttt{Cls-I} and \texttt{Cls-II} is used to fit the UHECR spectrum and composition for fixed values of $f_{\rm H}$. We vary $f_{\rm H}$ from $1.0-20.0$\%, at intervals of 0.5\% between $1.0-2.5$\% and at intervals of 2.5\% between $2.5-20.0$\% to save computation time. $\alpha_1$ is varied through the values 2.2, 2.4, and 2.6, inspired by previous analyses with light elements fitting the UHECR spectrum \citep{Aloisio_2017, Das_2019}. We vary $\log_{10}(R_{\rm cut,1}/\rm V)$ between the interval [19.5, 20.2] at grid spacings of 0.1, and $\log_{10}(R_{\rm cut,2}/\rm V)$ between [18.22, 18.36] at grid spacing of 0.02. For each combination of \{$\alpha_1, f_{\rm H}$\}, we find the best-fit values of $\log_{10}(R_{\rm cut,1}/\rm V)$, $\log_{10}(R_{\rm cut,2}/\rm V)$, $\alpha_2$, and composition $K_{i}$ at injection of \texttt{Cls-II}; that minimizes the $\chi^2_{\rm tot}$ of the combined fit. Due to increased number of parameters, we set the precision of composition $K_{i}$ to 0.25\%. These parameter sets are listed in Table~\ref{tab:two-pop}. For $\alpha_1=2.2$ and 2.4, the $\chi^2_{\rm tot}$ value monotonically increases with $f_{\rm H}$ beyond the best-fit value, while for $\alpha_1=2.6$, an alternating behaviour is obtained. The best-fits are found at $f_{\rm H}=$ 1.5\%, 2.5\%, and 2.0\%, respectively for $\alpha_1=2.2$, 2.4, and 2.6. For all the cases, a significant improvement in the combined fit is evident compared to the one-population model. It is worth pointing out that the minimum of $\chi^2_{\rm comp}$ and $\chi^2_{\rm spec}$ do not occur simultaneously and the variation in the best-fit value of $\log_{10}(R_{\rm cut,2}/\rm V)$ is insignificant. In the top right and bottom panels of Fig.~\ref{fig:spec_CTG}, we show the best-fit cases \rom{2}, \rom{14}, \rom{25} corresponding to $\alpha_1=2.2$, 2.4, and 2.6, respectively. The minimum $\chi^2$ value for all the three cases are comparable and very close to each other, indicating the best-fits are equally good for all the $\alpha_1$ values considered. The pure-proton component favors higher values of cutoff rigidity than \texttt{Cls-II} and steeper injection spectral index.

It is instructive to compare the all-flavor neutrino fluxes resulting from the two-population model with the current 90\% C.L. differential flux upper limits imposed by 9-years of IceCube data \citep{Aartsen_2018}. The hard spectral index and lower maximum rigidity in case of one-population model leads to a neutrino spectrum much lower than the current and upcoming neutrino detector sensitivities. This is shown in the top left panel of Fig.~\ref{fig:neu_CTG} along with the current sensitivity by PAO \citep{augerneu1, augerneu2} and that predicted for 3-years of observation by GRAND \citep{grand1, grand2} and POEMMA \citep{poemma1, poemma2}. We also present the allowed range of neutrino flux from \texttt{Cls-I} and \texttt{Cls-II} in the two-population model for $f_{\rm H}=1.0-20.0$\%. The cosmogenic neutrino flux from \texttt{Cls-I} is within the reach of the proposed GRAND sensitivity. The all-flavor integral limit for GRAND implies an expected detection of $\sim100$ neutrino events within 3-years of observation for a flux of $\sim10^{-8}$ GeV cm$^{-2}$ s$^{-1}$ sr$^{-1}$. This implies that with a further increase in exposure time, GRAND should be able to constrain our two-population model parameters if $f_H \gtrsim 10\%$.

\begin{figure}[hbt]
\centering
\includegraphics[width = 0.49\textwidth]{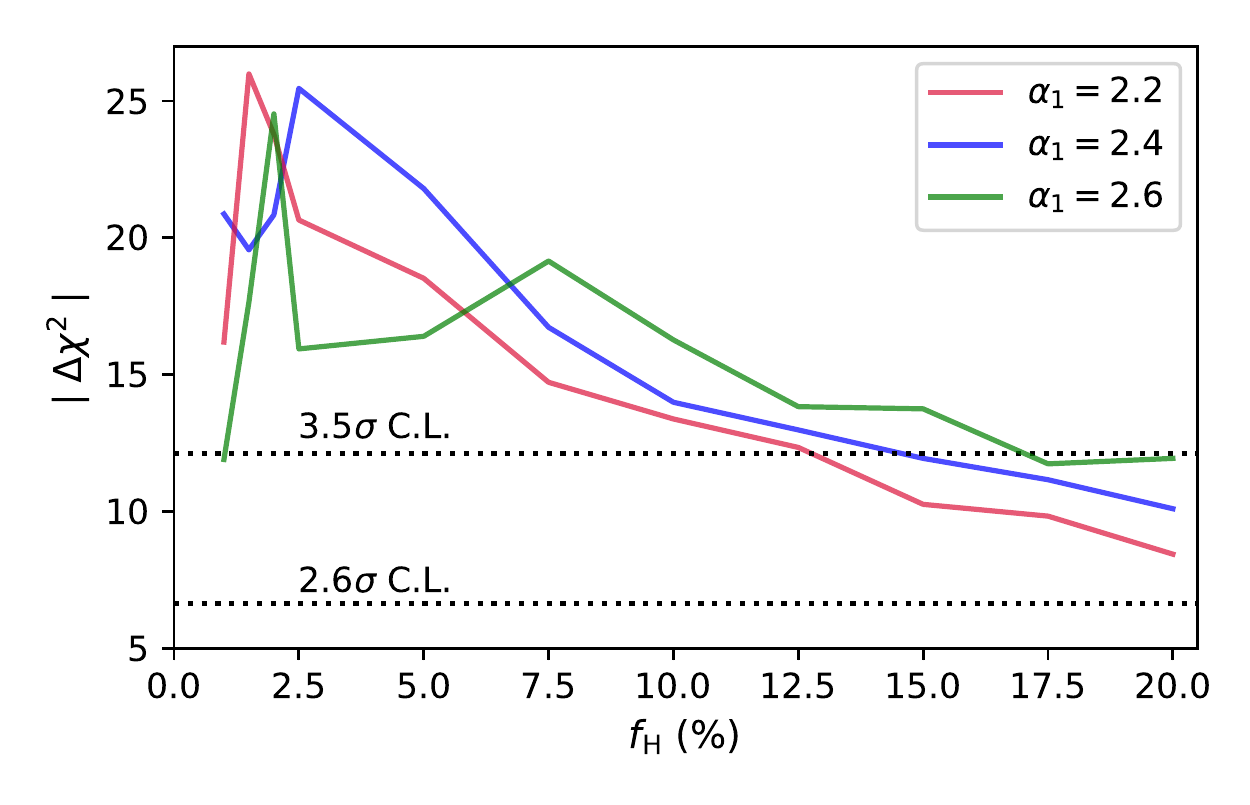}
\caption{\small{$\mid\Delta\chi^2\mid$ values between the one-population and two-population model (without cosmological evolutions) for one d.o.f are shown as a function of the pure-proton fraction $f_{\rm H}$. Three lines correspond to three values of \texttt{Cls-I} injection spectral index.}}
\label{fig:proton_frac}
\end{figure}

As we find the best-fit H fraction is zero in Table-I, $K_{\rm H}$ is a redundant parameter in this case. Scanning the parameter space excluding the latter will result in the same values of the remaining 6 parameters and thus, the resulting model coincides with that of \texttt{Cls-II} in Table-II. Thus, for a $\Delta\chi^2$ calculation between the one-population and two-population model, we consider the number of parameters in the former to be 6 and not 7. The difference in the number of parameters varied between one-population and two-population model is one, i.e., $R_{\rm cut,1}$. A smooth transition from the two-population model to one-population model can be done by setting $R_{\rm cut,1} =0$. This necessarily implies that $f_{\rm H}=0$ and there remains no $\alpha_1$. Based on the values obtained from,
\begin{equation}
\Delta\chi^2 = \chi^2\mid_{R_{cut,1}} - \chi^2\mid_{R_{cut,1}=0}
\end{equation} 
we estimate the maximum allowed proton fraction at $3.5\sigma$ confidence level (C.L.) in the highest-energy bin. For $\alpha_1=2.2$ this corresponds to $\approx12.5$\%, $\alpha_1=2.4$ corresponds to $\approx15.0$\%, and for $\alpha_1=2.6$ it turns out to be $\approx17.5$\%. However the maximum $\mid\Delta\chi^2\mid$, which also indicates the most significant improvement in contrast to one-population model, is found for $\alpha_1=2.2$, as shown in Fig.~\ref{fig:proton_frac}. The $2.6\sigma$ and $3.5\sigma$ C.L. are also indicated.

\subsection{\label{subsec:z_evol}Redshift evolution of sources}

\begin{figure*}
\centering
\begin{subfigure}{.5\textwidth}
  \centering
  \includegraphics[width=\textwidth]{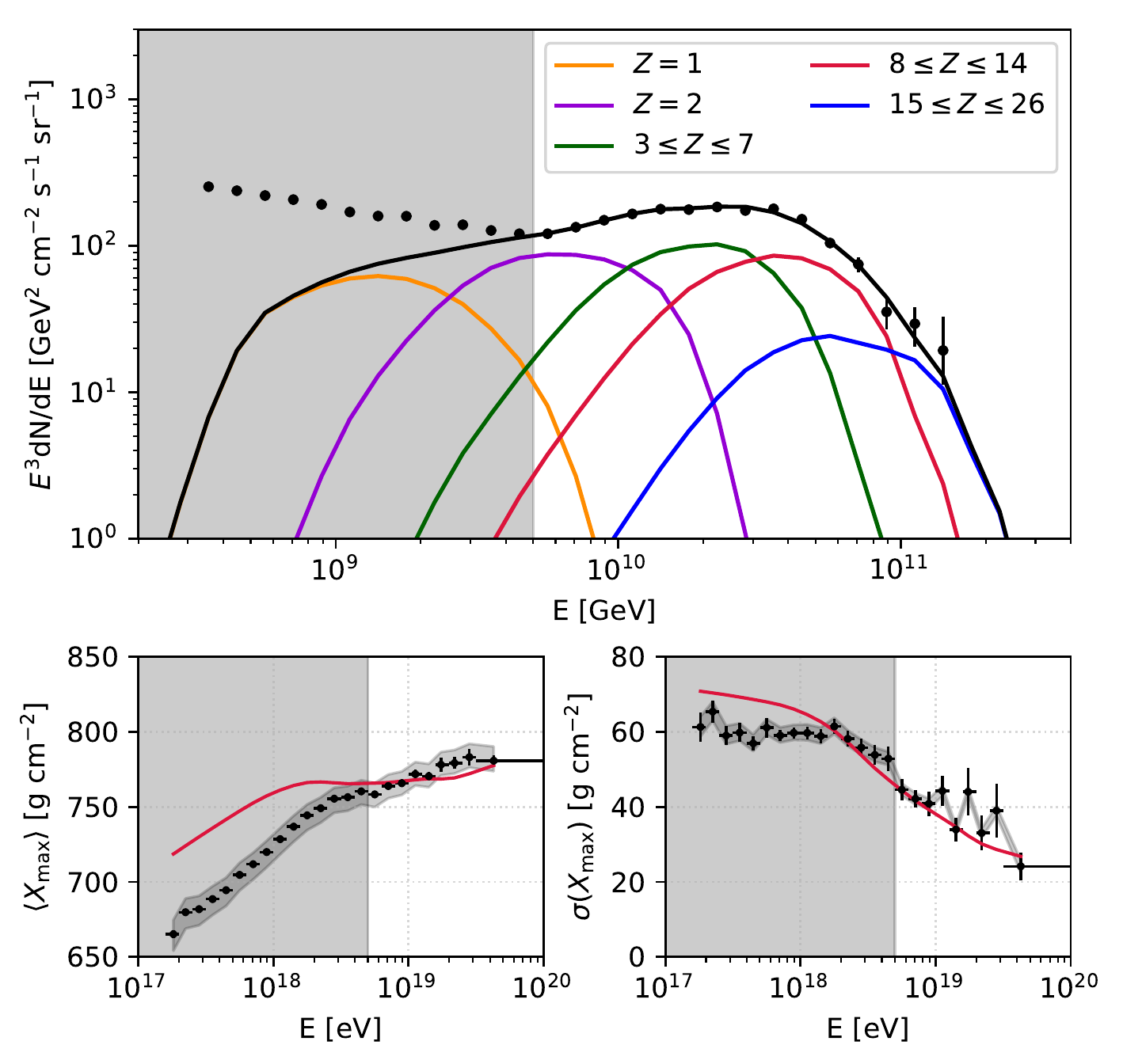}
\end{subfigure}%
\begin{subfigure}{.5\textwidth}
  \centering
  \includegraphics[width=\textwidth]{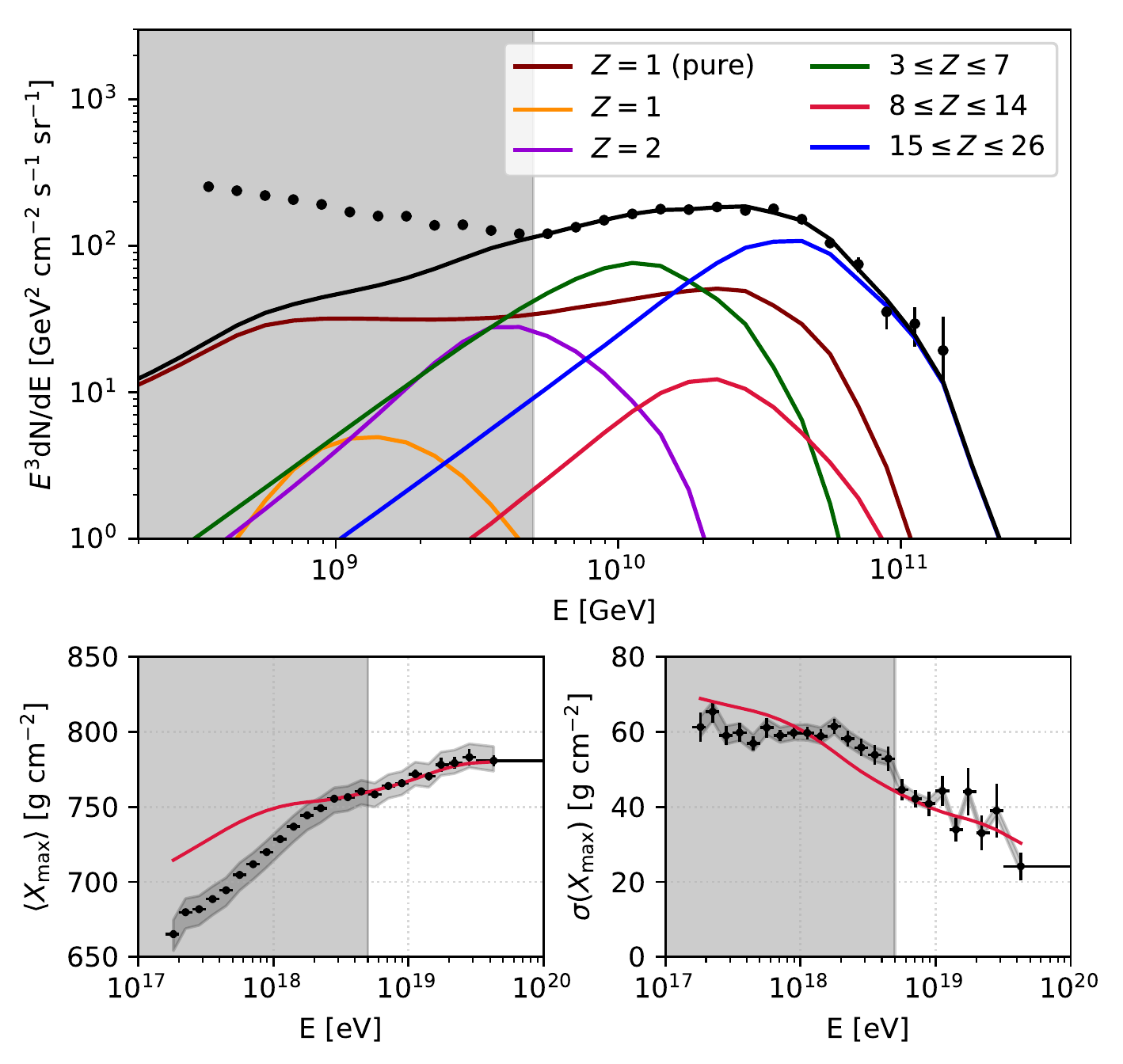}
\end{subfigure}
\begin{subfigure}{.5\textwidth}
  \centering
  \includegraphics[width=\textwidth]{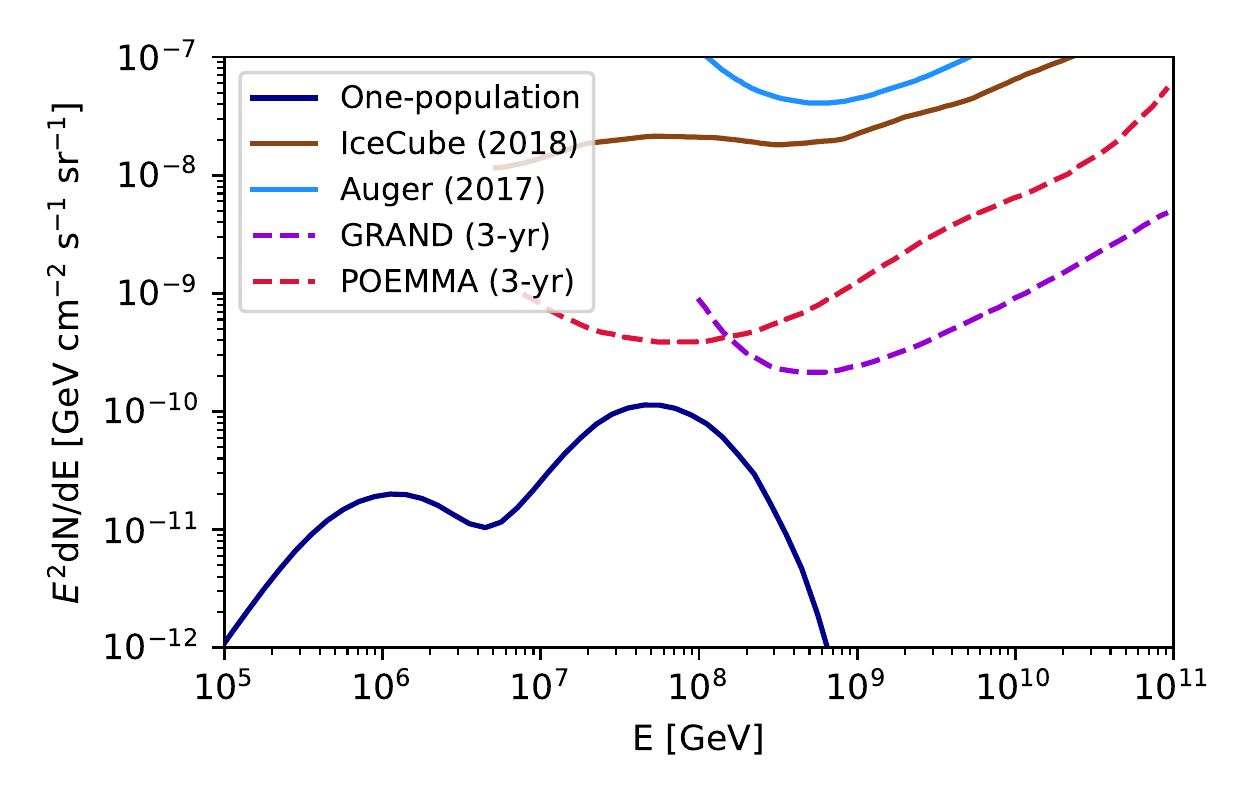}
\end{subfigure}%
\begin{subfigure}{.5\textwidth}
  \centering
  \includegraphics[width=\textwidth]{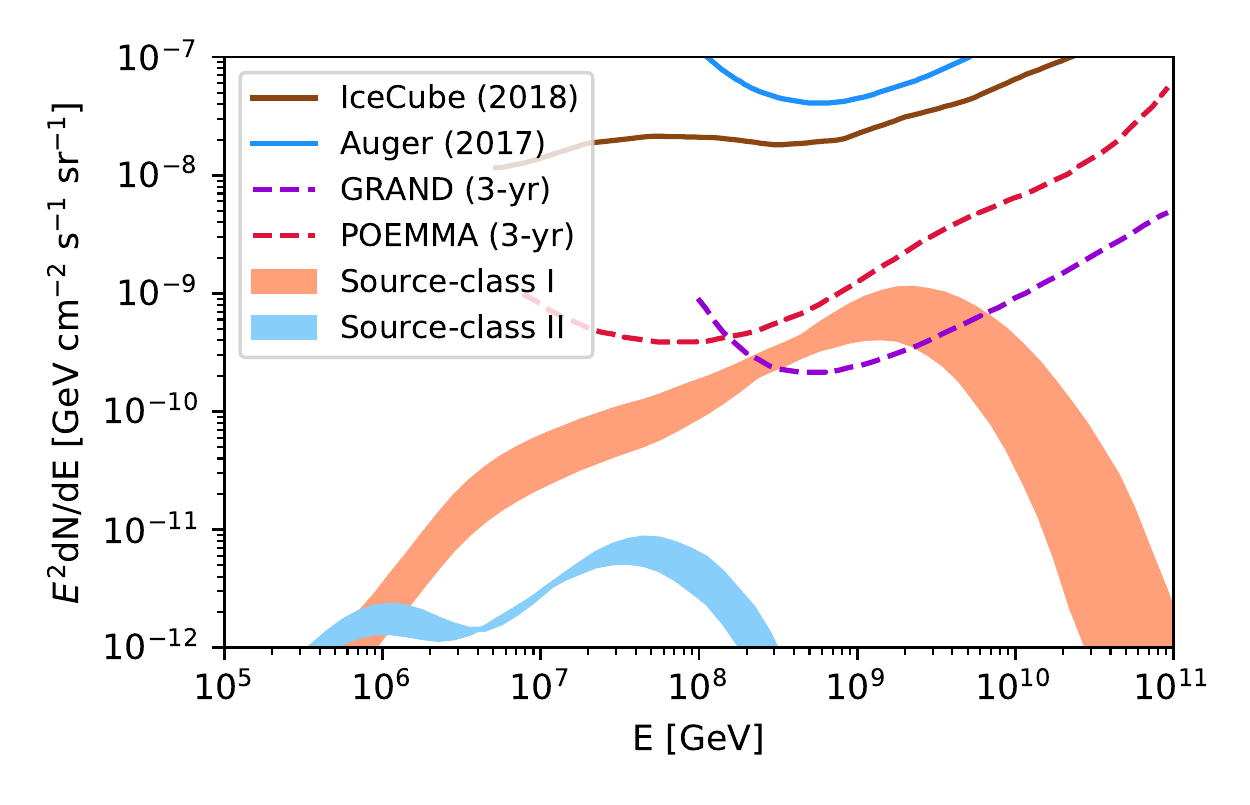}
\end{subfigure}
\caption{\small{The figures in the top panel indicates best-fit spectrum and composition obtained by considering redshift evolution as a free parameter, for one-population (\textit{left}) and two-population (\textit{right}) model, respectively. The lower panel indicates the corresponding neutrino fluxes. The shaded region (\textit{bottom right}) encompasses the neutrino spectrum possible for proton fraction $\sim$1.0\% to 20\% of the flux at the highest-energy bin of UHECR spectrum.}}
\label{fig:CF_evol}
\end{figure*}

In the preceding study with flat redshift evolution of the two populations of extragalactic sources, we see that the contribution of $^1$H from the light-to-heavy nuclei injecting sources, to the combined fit of energy spectrum and mass composition beyond the ankle, is infinitesimal. Whereas, the pure proton spectrum from \texttt{Cls-I} maintains a steady contribution up to the GZK energies superposed on the Peters cycle pattern \citep{Peters_1961}, resulting from \texttt{Cls-II}. We carry out a systematic analysis over plausible strengths of redshift evolution of the source classes. We assume the source distribution evolves with redshift according to $(1+z)^m$, where $m$ is a free parameter. First, we find out the best-fit value of $m$ in the one-population model assuming a mixed composition at injection comprising of $^4$He, $^{14}$N, $^{28}$Si, and $^{56}$Fe. The combined fit improves with comparison to the flat evolution case, but not substantially. The resulting spectrum and composition fit are shown in the top left panel of Fig.~\ref{fig:CF_evol}. The composition fit is found to be more significant than the flat evolution case. In this case too, we see the contribution from protons, resulting in the photodisintegration of heavier elements, is sub-dominant at $E\gtrsim10^{18.7}$ eV. The redshift evolution index $m$ is varied in the range $-6\leqslant m\leqslant+6$ at intervals of 1.0 and the corresponding best-fit values of cutoff rigidity, injection spectral index, and the composition are calculated. The best-fit parameters and the fit statistics are indicated in Table~\ref{tab:OP_evol}. The number of d.o.f is 25. The minimum $\chi^2$ is obtained for the fit corresponding to $m=+2$. This indicates a wide range of candidate classes, eg., low-luminosity GRBs (LL GRBs) where the UHECR nuclear survival is possible inside the source/jet \citep{Zhang_2018}. However, their redshift evolution is not well known but expected to follow that of long GRBs, given by $\psi(z)\propto(1+z)^{2.1}$ for $0<z<3$ \citep{Wanderman_2010}.

For the two-population case, we need to take into account two values of the redshift evolution index $m_1$ and $m_2$, respectively for \texttt{Cls-I} and \texttt{Cls-II}. In case of high-luminosity $\gamma$-ray sources, the dynamical timescales are larger than nuclear interaction timescales, inside the acceleration region. The relativistic jet provides suitable environment for heavier nuclei to dissociate via interactions with ambient matter and radiation. Hence, they are ideal candidate for $^1$H injection. We identify our \texttt{Cls-I} with AGNs, injecting predominantly protons. The redshift evolution of AGNs follow the function $\psi(z)\propto(1+z)^{3.4}$ for $z<1.2$ and X-ray luminosity in the range $L_X \sim 10^{43}-10^{44}$ erg/s \citep{Hasinger_2005}. An even higher luminosity might be required to accelerate UHECR protons up to $10^{20}$ eV \citep{Waxman_2004}. In principle, one can consider even higher luminosity AGNs, but the number density decreases sharply with luminosity. The redshift evolution of medium-high luminosity AGNs ($L_X \sim 10^{44}-10^{45}$) is given by $\psi(z)\propto (1+z)^{5.0}$ for $z<1.7$. Radio-loud quasers with bolometric $\gamma$-ray luminosity $10^{47}$ erg/s and a number density of $10^{-5}-10^{-4}$ Mpc$^{-3}$ can meet the energy requirements for UHECR acceleration \citep{Jiang_2007}. Hence, we vary $m_1$ through 3, 4, and 5. While for $m_2$, we consider a wide range of values spanning from positive to negative, viz., $+2$, 0, $-3$, and $-6$, to find the best-fit region. Once again, we fix the injection spectral index of proton-injecting sources to $\alpha_1=2.2$, 2.4, and 2.6, which is now a more physically motivated choice for AGNs. As before, we vary the proton fraction ($f_{\rm H}$) at the highest energy bin from $1.0-2.0\%$ at intervals of 0.5\%, and from $2.5\%-10.0\%$ at intervals of 2.5\%. We find the best-fit value of $R_{\rm cut,1}$, $R_{\rm cut,2}$, $\alpha_2$, and the fractional abundance of elements at injection ($K_i$) for \texttt{Cls-II}. For this case, we vary $\log_{10}(R_{\rm cut,2}/\rm V)$ with a precision of 0.1.

\begin{figure}
\centering
\begin{subfigure}{.48\textwidth}
\caption{Combined fit}
\includegraphics[width =\textwidth]{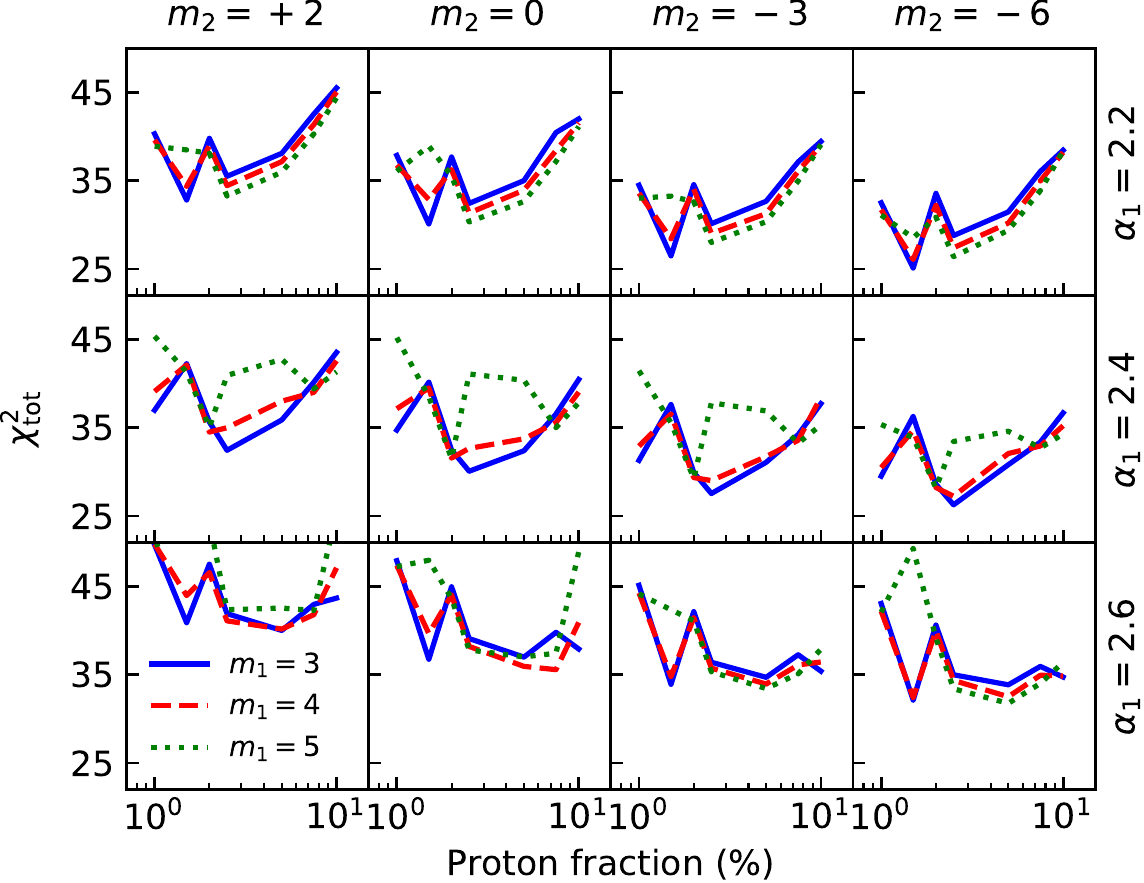}
\end{subfigure}
\begin{subfigure}{.48\textwidth}
\caption{Spectrum fit}
\includegraphics[width = \textwidth]{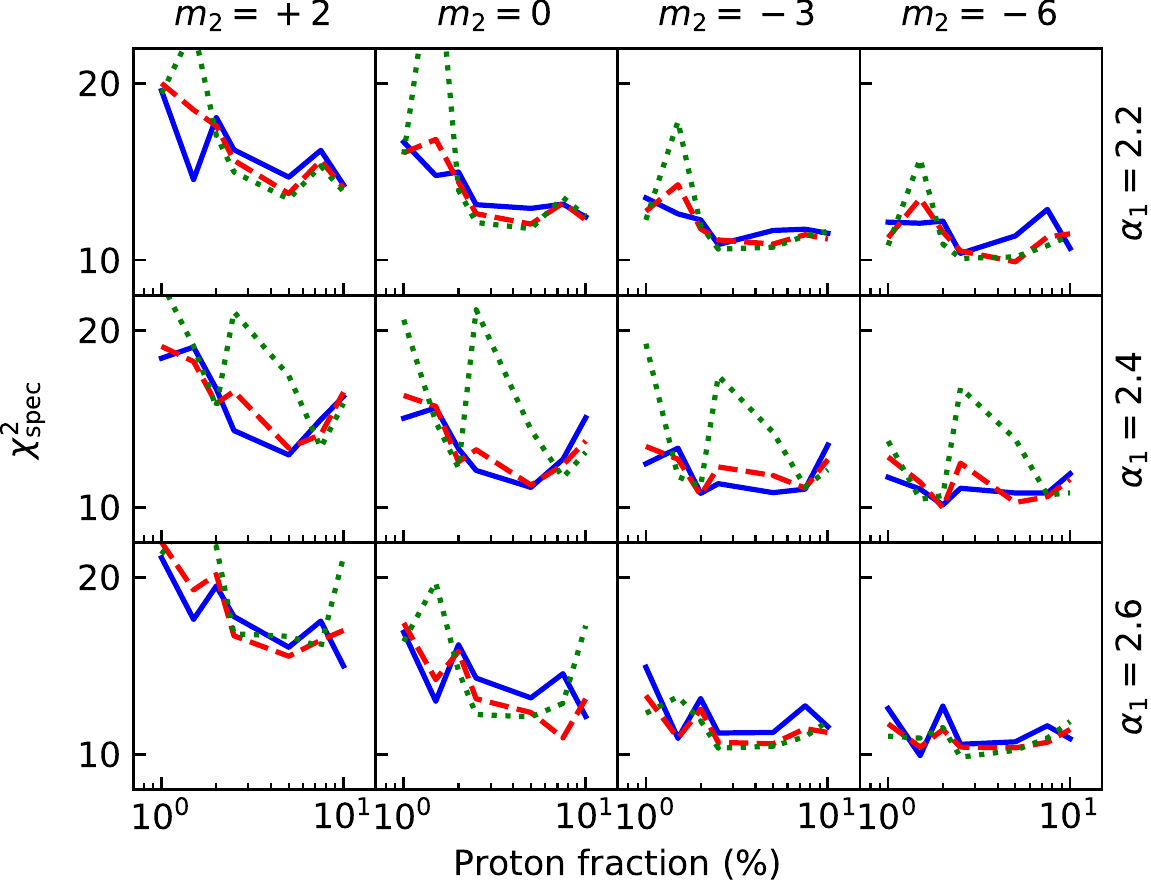}
\end{subfigure}
\begin{subfigure}{.48\textwidth}
\caption{Composition fit}
\includegraphics[width =\textwidth]{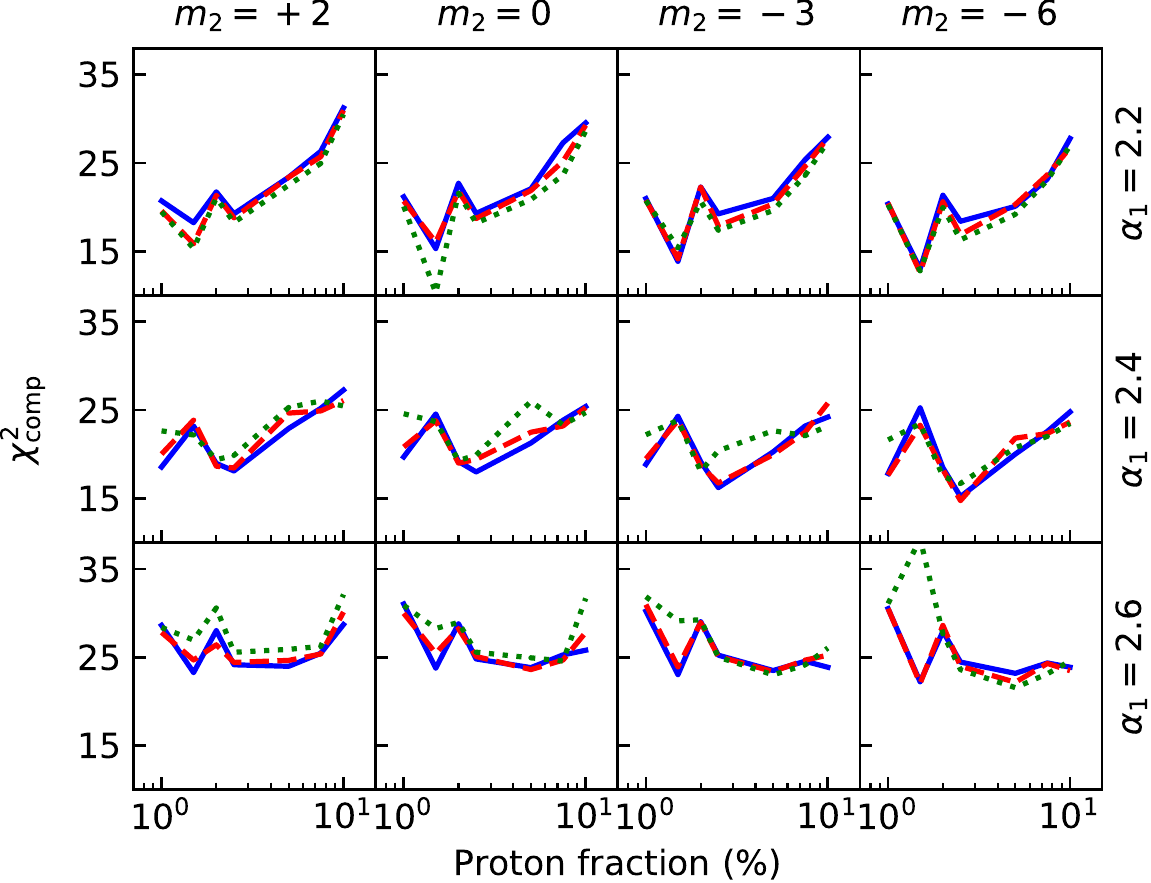}
\end{subfigure}
\caption{\small{$\chi^2$ values of the parameter scan for various values of $m_1$, $m_2$, $\alpha_1$, as a function of the proton fraction $f_{\rm H}$. The top, middle, and bottom panels indicate the values for combined fit, spectrum fit, and composition fit, respectively.}}
\label{fig:redshift_space}
\end{figure}

\begin{table}
\caption{\label{tab:OP_evol}UHECR best-fit parameter set for the one-population model} 
\centering
\begin{tabular}{lllll}
\hline
\textbf{Parameter} & \multicolumn{3}{l}{\textbf{Description}} & \textbf{Values}\\ 
\hline
$\alpha$ & \multicolumn{3}{l}{Source spectral index} & -0.9 \\ 
$\log_{10}(R_{\rm cut}/\rm V)$ & \multicolumn{3}{l}{Cutoff rigidity} & 18.2 EV\\
$z_{\text{max}}$ & \multicolumn{3}{l}{Cutoff redshift} & 1.0 (fixed)\\
$m$ & \multicolumn{3}{l}{Source evolution index} & +2.0 \\
\hline
$K_i$(\%) & He & \hspace*{0.3cm} N & \hspace*{0.4cm} Si \hspace*{0.8cm} Fe\\
 &  94.4 & \hspace*{0.3cm} 5.3 & \hspace*{0.4cm} 0.3 \hspace*{0.5cm} 0.007\\
 \hline
 $\chi^2_{\rm tot}/$d.o.f & $\chi^2_{\rm spec}$ & $\chi^2_{\rm comp}$ & &\\
 52.43/25 & 11.66 & 40.77 & &\\
 \hline%
\end{tabular}
\end{table}

We represent the goodness-of-fit for the best-fit case corresponding to each set of \{$\alpha_1$, $f_{\rm H}$, $m_1$, $m_2$\} values in Fig.~\ref{fig:redshift_space}, distinctly for the combined fit, spectrum fit, and the composition fit from top to bottom, respectively. The combined fit improves as we go to more and more negative values of $m_2$ and lower values of $\alpha_1$. We see the best-fit occurs for $m_1=+3$, $m_2=-6$, $\alpha_1=2.2$ and $f_{\rm H}=1.5\%$. The details for this set are given in Table~\ref{tab:TP_evol}. The best-fit spectrum and composition are displayed on the top right panel of Fig.~\ref{fig:CF_evol}. However, it is interesting to note from Fig.~\ref{fig:redshift_space} that the best-fit composition and best-fit spectrum cases are not coincident. The best-fit composition ($\chi^2_{\rm comp}= 10.29$) is obtained for $m_1=+5$, $m_2=0$, $\alpha_1=2.2$ and $f_{\rm H}=1.5\%$, whereas the best-fit spectrum ($\chi^2_{\rm spec}$) occurs at $m_1=+5$, $m_2=-6$, $\alpha_1=2.6$ and $f_{\rm H}=2.5\%$. We also calculate the neutrino fluxes originitaing from the best-fit one-population and two-population models, after considering redshift evolution of the source classes. In the bottom left panel of Fig.~\ref{fig:CF_evol}, the neutrino flux increases in the case of one-population model, owing to the positive redshift evolution ($m=2$). While, in case of two-population model, the flux from heavy nuclei injecting sources is greatly reduced ($m_2=-6$), and the cumulative neutrino flux distribution is dominated by that from protons ($m_1=3$). The shaded region indicates the flux range enclosed by $f_{\rm H}=1.0-20.0\%$, $\alpha_1=2.2$, and is within the flux upper limit imposed by 9-yr of IceCube data. Even a small fraction of proton can yield a neutrino flux which is within the reach of 3-yr extrapolated sensitivity of the proposed GRAND detector.

The preference over large negative values of $m_2$ can be attributed to specific source classes, such as tidal disruption events (TDEs) \citep{Farrar_2009, Farrar_2014}. The event rate of TDEs depend on the number density of SMBH as a function of redshift. The best-fit empirical model indicates a negative redshift evolution \citep{Sun_2015}. TDEs forming relativistic jets can be the powerhouse of UHECR acceleration, but their event rate severely constrains the UHECR flux \citep{Murase_2008a}, thus requiring a mixed or heavy composition at injection. Metal-rich composition consisting of a significant Si and Fe fraction is required to explain the spectrum with a population of TDE \citep{Guepin_2018}. In our case too, a high fraction of Fe is required to explain the spectrum at the highest energies, as indicated in Table~\ref{tab:TP_evol}. However, the survival of UHECR nuclei depends on the specific outflow model. For luminous jetted TDEs like Swift J1644+57 \citep{Bloom_2011, Burrows_2011}, which reaches a bolometric luminosity $L_{\rm bol}\gtrsim10^{48}$ erg/s in the high state, UHECR acceleration becomes difficult via internal shock model, but is allowed for TDEs with lower luminosities. Forward/reverse shock models were also found in accordance with heavy nuclei injection \citep{Zhang_2017}.

\begin{table*}[hbt]
 \caption{\label{tab:TP_evol} Best-fits to UHECR spectrum and composition for two-population model ($m_1=+3$, $m_2=-6$)}
  \begin{ruledtabular}
\centering
 \begin{tabular}{ccc|cccccc|ccc|l}
\multicolumn{3}{c|}{\textbf{\texttt{Source-class I}}} & \multicolumn{6}{c|}{\textbf{\texttt{Source-class II}}} & \multicolumn{4}{c}{\textbf{Goodness-of-fit}}\\ [0.15cm]
\hline
$\alpha_{1}$ & $f_{\rm H}$(\%) & $\log_{10}(R_{\rm cut, 1}/\rm V)$ & $\alpha_{2}$ & $\log_{10}(R_{\rm cut, 2}/\rm V)$ & $K_{He}$ & $K_{N}$ & $K_{Si}$ & $K_{Fe}$ & $\chi^2_{\rm spec}$ & $\chi^2_{\rm comp}$ & $\chi^2_{\rm tot}$ & ID \#\\ [0.15cm]
\hline
\textbf{2.2} & 1.5\% & 19.5 & 1.6 & 18.3 & 33.00 & 54.75 & 2.25 & 10.00 & 12.09 & 13.00 & 25.09 & SRE-1\\
\textbf{2.4} & 2.5\% & 19.6 & 1.6 & 18.3 & 23.50 & 62.00 & 3.75 & 10.75 & 11.07 & 15.19 & 26.26 & SRE-2\\
\textbf{2.6} & 1.5\% & 19.6 & 1.5 & 18.3 & 24.75 & 60.75 & 5.50 & 9.00 & 9.92 & 22.20 & 32.12 & SRE-3\\
 \end{tabular}
 \end{ruledtabular}
\end{table*}

\begin{figure}
\centering
\includegraphics[width = 0.49\textwidth]{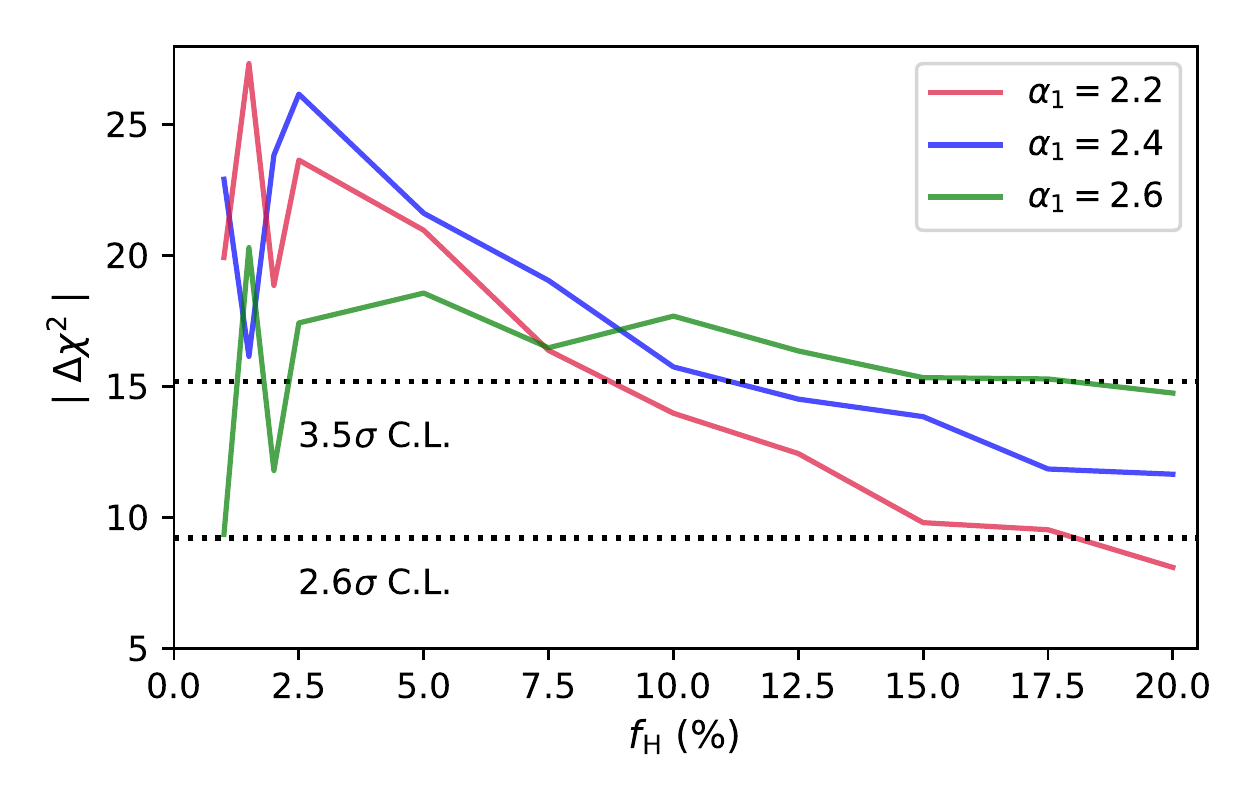}
\caption{\small{$\mid\Delta\chi^2\mid$ values between the one-population and two-population model for non-trivial redshift evolution and two d.o.f as a function of the pure-proton fraction $f_{\rm H}$. Three lines correspond to three values of proton injection index.}}
\label{fig:PF_evol}
\end{figure}

\begin{figure*}
\centering
\includegraphics[width = 0.8\textwidth]{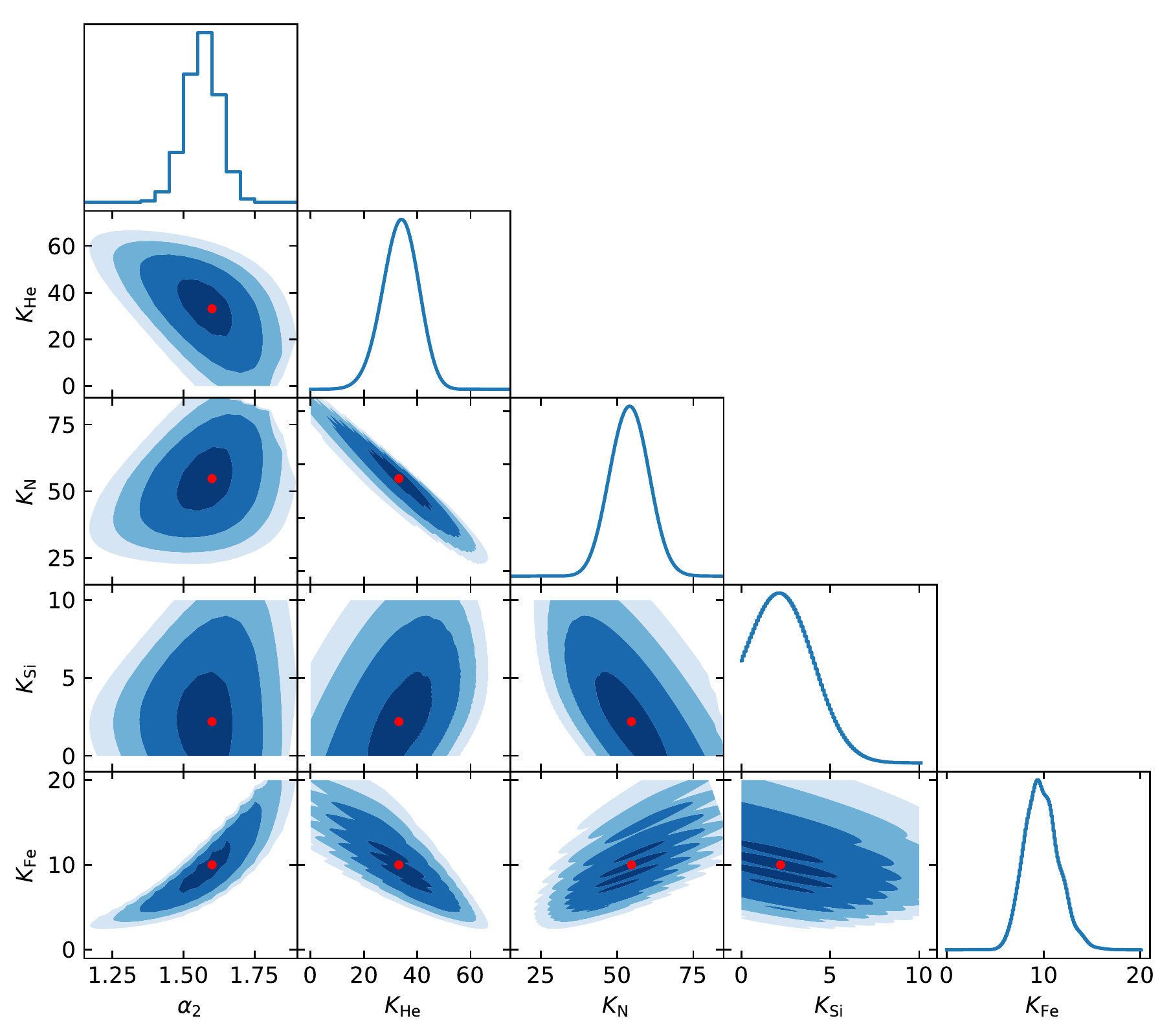}
\caption{\small{Correlation between fit parameters for the best-fit case corresponding to $m_1=3$, $m_2=-6$, $\alpha_1=2.2$, and $f_{\rm H}=1.5$\%, i.e., for the best-fit case SRE-1 listed in Table~\ref{tab:TP_evol}. It can be seen that a high fraction of Fe is required at injection along with a hard injection spectral index. The diagonal plots represent the posterior probability distribution and red dots in others indicate the central values. The $1\sigma$, $2\sigma$, and $3\sigma$ standard deviations are shown by dark to light-colored shading.}}
\label{fig:fit_corr}
\end{figure*}

The best-fit obtained in two-population model with non-trivial redshift evolution ($m_1$, $m_2\neq$0) is better than the flat evolution case and also compared to the one-population model with redshift evolution. A smooth transition can be made from two-population model to one-population model by setting $R_{\rm cut,1}=0$ and $m_1=0$. So the difference in the number of parameters varied is two. Fig.~\ref{fig:PF_evol} shows the $\mid\Delta\chi^2\mid$ values for two d.o.f between the one-population and two-population cases, as a function of proton fraction at the highest-energy bin. As before, we constrain the maximum allowed proton fraction at 3.5$\sigma$ confidence level, which turns out to be between 7.5\% and 15\% for different values of the proton injection spectral index considered. We also calculate the correlation between the fit parameters for the specific case of $m_1=3$, $m_2=-6$, $\alpha_1=2.2$, and $f_{\rm H}=1.5$\%, i.e., for the best-fit case SRE-1 listed in Table~\ref{tab:TP_evol}. We vary the cutoff rigidity $\log_{10}(R_{\rm cut,1}/\rm V)$ in the range 19.4 to 20.0 at intervals of 0.1, and $\log_{10}(R_{\rm cut,2}/\rm V)$ between 18.1 and 18.5 with grid spacing of 0.1. The \texttt{Cls-II} injection spectral index $\alpha_2$ is varied over the range 1.0 to 2.0 at intervals of 0.05. The composition fractions $K_i$ are varied at intervals of 0.25\%. The number of d.o.f are $33-7-1=25$, where we consider the normalization to be an additional free parameter. Fig.~\ref{fig:fit_corr} shows the $1\sigma$, $2\sigma$, and $3\sigma$ C.L. contours for 25 d.o.f. The variation of cutoff rigidity is not shown, since they were found to be insensitive to variation of other parameters. We see a hard injection spectral index $\alpha_2\approx1.6$ is preferred, which conforms with the recent predictions by the PAO \citep{PAO_2017JCAP}, and analysis done by other works \citep{Batista_2019, Heinze_2019}. The composition is also in accordance with those predicted from latest measurements, implying a progressively heavier composition at higher energies. The $1\sigma$ region in composition space corresponds to a high value of Fe fraction, which is indeed needed for TDEs to have significant contribution in the UHECR spectrum. Another candidate class which can represent our \texttt{Cls-II} is the low-luminosity ($L_\gamma<10^{44}$ erg/s) and high synchrotron peaked BL Lacertae objects. They possess a negative redshift evolution and are predicted to be more numerous than their high-luminosity counterpart. However, a direct detection of these low-luminosity objects are difficult, and the current 4LAC catalog consists $\sim20$ such sources.

\section{\label{sec:discussions}Discussions \protect}

The composition fit corresponding to the one-population model, especially the departure of simulated $\langle X_{\rm max}\rangle$ and $\sigma (X_{\rm max})$ values from the data, leaves a substantial window for improvement. The addition of a light nuclei component up to the highest observed energies shall alleviate the mismatch. We exploit this possibility in our work by adding a distinct source population injecting $^1$H that extends up to the highest observed energies. Earlier works have considered a pure-protonic component with an assumed steep injection spectral index \citep{Aloisio_2014} to explain the region of the spectrum below the ankle. We do not fit the UHECR spectrum below the ankle, and the proton spectrum considered in our work contributes directly to the improvement in composition fit at the highest energies. A relatively hard spectrum ($\alpha_1=-1$) in addition to a Milky Way-like nuclear composition is considered in Ref.~\citep{Muzio_2019}, extending up to the highest energies. We do not assume any fixed abundance fraction for the light-to-heavy nuclei injecting sources and calculate the best-fit values within the resolution adopted.

Ref.~\citep{Vliet_2019} have proposed an interaction-model independent method to probe the allowed proton fraction for $E_p\gtrsim 30$ EeV, constrained by the cosmogenic neutrino flux upper limits at 1 EeV. Thus, they do not take the composition of primary cosmic rays into account, inferred from air shower data. They have considered a generalized redshift evolution function of the proton injecting sources, parametrized by the evolution index $m$. In our work, we fit the composition data $X_{\rm max}$, $\sigma(X_{\rm max})$, and the energy spectrum simultaneously to infer the proton fraction in a two-population scenario.

Here, we find a significant improvement in the combined fit to spectrum and composition data, when adding an extragalactic source population emitting UHECRs as protons. For our choice of steep proton injection indices ($\alpha_1$), the goodness-of-fit is found to be comparable to each other. We also consider the injection index ($\alpha_2$), maximum rigidity ($R_{\rm cut,2}$), and composition fractions ($K_i$) of the second population injecting light-to-heavy nuclei to be variables and find the corresponding best-fit values. The corresponding improvement in the combined fit is found to be $\gtrsim 3\sigma$ in some cases. 

We have also surveyed our results for a wide range of source redshift evolution. Such an analysis is already done earlier for a single source population for a mixed composition of injected elements \citep{Batista_2019, Heinze_2019, Taylor_2015}. In our analysis, we find that, although a positive evolution index is preferred in one-population model, the best-fit value changes sign on going to two-population model. However, with increasing values of $z_{\rm max}$, the variation of $m$ can significantly affect the neutrino spectrum. We have kept the contributing sources within $z\lesssim1$ in view of the fact that particles originating at higher redshifts will contribute below the ankle, which we do not fit here. Thus within the minimal requirements of this model, our neutrino spectrum can be considered as a conservative lower bound in the two-population scenario.

The resultant neutrino spectrum in two-population model at $E\gtrsim0.1$ EeV is dominated by that from pure-protons. Even a small fraction of protons at the highest energy is capable of producing a significant flux of neutrinos. This is expected because of the maximum energy considered for proton-injecting sources. Even for low $f_{\rm H}$, the values of $E_{\rm max}$ are very close to GZK cutoff energy, where the resonant photopion production occurs, leading to pion-decay neutrinos. The double-humped feature of the neutrino spectrum is a signature of interactions on the CMB and EBL by cosmic rays of different energies. The higher energy peak produced from protons possesses the highest flux, and the detection of these neutrinos at $\sim3\cdot10^{18}$ eV will be a robust test of the presence of a light component at the highest energies, thus also constraining the proton fraction. For $E<0.1$ EeV, the neutrinos from \texttt{Cls-II} becomes important with peaks at $\sim1$ PeV and $\sim40$ PeV. Hence, the cumulative neutrino spectrum (\texttt{Cls-I} + \texttt{Cls-II}) exhibits three bumps for $\alpha_1=2.2$ (see Fig.~\ref{fig:neu_CTG}). But gradually with increasing values of $\alpha_1$, the lower energy peak of \texttt{Cls-I} becomes significant, diminishing the ``three-peak" feature until neutrinos from protons dominate down to $\sim1$ PeV for $\alpha_1=2.6$

We present the upper limit on the maximum allowed proton fraction in two-population model at $\approx1.4\times10^{20}$ eV. This is based on the improvement in the combined fit compared to the one-population model, up to $3.5\sigma$ statistical significance. For a higher C.L., the proton fraction is even lower at the highest-energy bin. However, a non-zero proton fraction is inevitable. It is studied earlier that the flux of secondary photons increases with an increasing value of $\alpha_1$ \citep{Berezinsky_2011}. If a single population injecting protons is used to fit the UHECR spectrum, the resulting cosmogenic photon spectrum saturates the diffuse gamma-ray background at $\sim1$ TeV for $\alpha_1=2.6$, $m=0$ \citep{Das_2019}. In our two-population model, the proton fraction at the highest energies is much lower than the total observed flux. This ensures the resulting photon spectrum from \texttt{Cls-I} is well within the upper bound imposed by Fermi-LAT \citep{Ackermann_2015}. For {\texttt{Cls-II}} injecting heavier nuclei, the main energy loss process is photodisintegration, contributing only weakly to the cosmogenic photon flux. Hence the two-population model, which we invoke in our study, is in accordance with the current multimessenger data.

The choice of the hadronic interaction model for our analysis is based on the interpretation of air shower data by the PAO \citep{Abreu_2013, PAO_ICRC_2017}. It is found that \textsc{qgsjet-II.04} is unsuitable compared to the other two models and leads to inconsistent interpretation of observed data \citep{PAO_comp2017}. Also, for our choice of photodisintegration cross-section, i.e. \textsc{talys} 1.8, the hadronic model \textsc{sybill2.3c} yields superior fits \citep{Heinze_2019}. In general, the \textsc{sybill2.3c} model allows for the addition of a higher fraction of heavy nuclei, compared to others, at the highest energies. Indeed in Table~\ref{tab:two-pop}, it is seen that the lowest-$\chi^2$ cases correspond to high $K_{\rm Fe}$, which increases monotonically with $\alpha_1$. The requirement of Fe abundance in one-population model is much lower than in the case of two-population model. For the latter, the cutoff in the cosmic ray spectrum cannot be solely explained by the maximum acceleration energy of iron nuclei at the sources, but also, must be attributed to photopion production of UHECR protons on the CMB to some extent. 

In going from one-population to the two-population model, the injection spectral index of the population injecting heavier elements changes sign from negative to positive, making it easier to accept in the context of various astrophysical source classes. Young neutron stars, eg., can accelerate UHECR nuclei with a flat spectrum, $\alpha_2 \sim1$ \citep{Blasi_2000}. Particle acceleration in magnetic reconnection sites can also result in such hard spectral indices \citep[see for eg.,][]{Kowal_2011}. Luminous AGNs and/or GRBs are probably candidates for \texttt{Cls-I}, accelerating protons to ultrahigh energies \citep{Waxman_1995}. The \texttt{Cls-II} injecting light-to-heavy nuclei suggests the sources to be compact objects or massive stars with prolonged evolution history, leading to rich, heavy nuclei abundance in them. In particular the high negative redshift evolution and substantial Fe fraction allows us to identify the \texttt{Cls-II} with TDEs. The problem in the case of a highly luminous object is, although heavier nuclei may be accelerated in the jet, they interact with ambient matter and radiation density in the environment near the sources \citep{Wang_2008}. To increase the survivability of UHECR nuclei, less luminous objects such as LL GRBs \citep{Murase_2008} are preferred.

\section{\label{sec:conclusions}Conclusions\protect} 

Based on the spectrum and composition data measured by PAO, a combined fit analysis with a single-population of extragalactic sources suggest that the composition fit at the highest energy deserves improvement. The slope of the simulated $\langle X_{\rm max}\rangle$ curve implies that fitting the highest-energy data points with contribution from only $^{56}$Fe will diminish the abundance of lighter components $^{28}$Si, $^{14}$N, and $^4$He. This will in turn decrease the flux near the ankle region, thus resulting in a bad fit. Addition of another light component of extragalactic origin, preferably pure proton, extending up to the highest-energy bin can resolve this problem. From a critical point of view, this solution is not unique, but definitely a rectifying one. The combined fit improves significantly and we present the maximum allowed proton fraction at the highest-energy bin of spectrum data corresponding to $>3\sigma$ statistical significance. An additional population of extragalactic protons has also been suggested in Ref.~\cite{Gaisser_2013}, in the context of fitting the UHECR spectrum.

There are observational indications that different astrophysical source populations likely contribute to the UHECR data.  A plausible hot spot around the nearby starburst galaxy M82 \citep{Abbasi_2014, Pfeffer_2016} in the TA data and an intermediate-scale anisotropy around the nearest radio galaxy Cen-A in the Auger data already suggest possibility of two types of source populations. The Auger data, however, do not show any small-scale anisotropy, suggesting that the majority of UHECR sources are distributed uniformly in the sky. The recent 3$\sigma$ correlation of an observed high-energy muon neutrino event detected by IceCube with a flaring blazar TXS 0506+056 at a moderate redshift of 0.34 is consistent with this scenario \cite{IceCube:2018cha, IceCube:2018dnn}.

The two generic source classes of UHECRs studied here by us is also representative of the scenario described above. High luminosity AGNs or GRBs could contribute a pure proton component that is significant at the highest UHECR energies. The resulting cosmogenic neutrino spectrum can be detected by future experiments with sufficient exposure and the proton fraction in the highest energy UHECR data can be tested.
\begin{acknowledgements}
We thank Antonella Castellina, Ralph Engel, Jose Bellido, Francesco Fenu, and the Pierre Auger Collaboration for useful correspondence regarding the spectrum and composition data. We also thank Sergio Petrera for providing the latest parametrizations of $X_{\rm max}$ moments obtained from \textsc{conex} simulations using various hadronic interaction models. S.D. thanks Nirupam Roy for useful discussions. The work of S.R. was partially supported by the National Research Foundation (South Africa) with grant No. 111749 (CPRR) and by the University of Johannesburg Research Council grant.
\end{acknowledgements}



\nocite{*}

\bibliography{two_source}

\end{document}